\documentclass[journal,10pt]{IEEEtran}
\usepackage{multirow}
\usepackage{tabu}
\usepackage{makecell}
\usepackage{color}
\usepackage[table]{xcolor}
\usepackage{graphicx}
\usepackage{bm}
\usepackage{balance}
\usepackage{amsmath,amssymb,amsfonts}

\usepackage{bbding}

\begin{document}

\title{MTH-IDS: A Multi-Tiered Hybrid Intrusion Detection System for Internet of Vehicles}


\author{\IEEEauthorblockN{Li Yang, Abdallah Moubayed, and Abdallah Shami}\\
\IEEEauthorblockA{
Western University, London, Ontario, Canada \\
e-mails: \{lyang339, amoubaye, abdallah.shami\}@uwo.ca}



}

\markboth{Accepted and to Appear in IEEE Internet of Things Journal}
{}

\maketitle

\begin{abstract}
Modern vehicles, including connected vehicles and autonomous vehicles, nowadays involve many electronic control units connected through intra-vehicle networks to implement various functionalities and perform actions. Modern vehicles are also connected to external networks through vehicle-to-everything technologies, enabling their communications with other vehicles, infrastructures, and smart devices. However, the improving functionality and connectivity of modern vehicles also increase their vulnerabilities to cyber-attacks targeting both intra-vehicle and external networks due to the large attack surfaces. To secure vehicular networks, many researchers have focused on developing intrusion detection systems (IDSs) that capitalize on machine learning methods to detect malicious cyber-attacks. In this paper, the vulnerabilities of intra-vehicle and external networks are discussed, and a multi-tiered hybrid IDS that incorporates a signature-based IDS and an anomaly-based IDS is proposed to detect both known and unknown attacks on vehicular networks. Experimental results illustrate that the proposed system can detect various types of known attacks with 99.99\% accuracy on the CAN-intrusion-dataset representing the intra-vehicle network data and 99.88\% accuracy on the CICIDS2017 dataset illustrating the external vehicular network data. For the zero-day attack detection, the proposed system achieves high F1-scores of 0.963 and 0.800 on the above two datasets, respectively. The average processing time of each data packet on a vehicle-level machine is less than 0.6 ms, which shows the feasibility of implementing the proposed system in real-time vehicle systems. This emphasizes the effectiveness and efficiency of the proposed IDS.
\end{abstract}


\IEEEpeerreviewmaketitle

\section{Introduction}
With the increasing research and rapid development of the Internet of Vehicles (IoV) technology, connected vehicles (CVs) and autonomous vehicles (AVs) are becoming increasingly popular in the modern world \cite{IoV}. IoV serves as a primary vehicular communication framework that enables reliable communications between vehicles and other IoV entities, such as infrastructures, pedestrians, and smart devices. \cite{IoV}.

IoV consists mainly of intra-vehicle networks (IVNs) and external vehicular networks \cite{IoV}. IVNs involve an increasing number of electronic control units (ECUs) to adopt various functionalities \cite{ECU}. All ECUs in a vehicle are connected by a controller area network (CAN) bus to transmit messages and perform actions \cite{CANrisk}. On the other hand, external networks connect modern vehicles to the outer environment by vehicle-to-everything (V2X) technologies. V2X technology allows modern vehicles to communicate with other vehicles, roadside infrastructures, and road users \cite{V2XCAN} \cite{thesis}.

However, with the increasing level of connectivity and complexity of modern vehicles, their security risks have become a significant concern. Cyber threats may decrease the stability and robustness of IoV, as well as cause vehicle unavailability or traffic accidents. A real-life example can be found in \cite{realeg}: two attackers compromised and fooled a jeep car into performing dangerous actions, including turning the steering wheel and activating the parking brake at highway speeds, causing severe accidents. In IVNs, CANs are mainly vulnerable to message injection attacks due to their broadcast communication strategy and the lack of authentication \cite{V2XCAN}. In external networks of IoV, vehicle systems are exposed to various common cyber-attacks, like denial-of-service (DoS), sniffing, and global positioning system (GPS) spoofing attacks \cite{globecom}. This is because, in large external vehicular networks comprising various types of networks and entities, every node is a potential entry point for cyber-attacks.  

Many traditional security mechanisms, like certain authentication and cryptographic techniques, are unsuitable for intra-vehicle networks because they are not supported in CANs or may violate timing constraints of CAN communications \cite{timing}. Thus, intrusion detection systems (IDSs) have become an essential component in modern IoV to identify malicious threats on vehicular networks \cite{GIDS}. IDSs are often incorporated into external networks as an essential component of the defense system to identify malicious attacks that can breach firewalls and authentication mechanisms. Although many previous works have made some success developing IDSs, intrusion detection is still a challenging problem due to the high volume of network traffic data, numerous available network features, and various cyber-attack patterns \cite{globecom}. 

Machine learning (ML) and data mining algorithms have been recognized as effective models to design IDSs \cite{MLreview_AM}. In this paper, a multi-tiered hybrid intrusion detection system (MTH-IDS) is proposed to efficiently identify known and zero-day cyber-attacks on both intra-vehicle and external networks using multiple ML algorithms. 
The proposed MTH-IDS framework consists of two traditional ML stages (data pre-processing and feature engineering) and four tiers of learning models: 
\begin{enumerate}
\item Four tree-based supervised learners — decision tree (DT), random forest (RF), extra trees (ET), and extreme gradient boosting (XGBoost) — used as multi-class classifiers for known attack detection; 
\item A stacking ensemble model and a Bayesian optimization with tree Parzen estimator (BO-TPE) method for supervised learner optimization; 
\item A cluster labeling (CL) k-means used as an unsupervised learner for zero-day attack detection; 
\item Two biased classifiers and a Bayesian optimization with Gaussian process (BO-GP) method for unsupervised learner optimization. 
\end{enumerate}

Therefore, tiers 1 and 3 of the MTH-IDS are designed for basic known and unknown attack detection functionalities, respectively, while tiers 2 and 4 are designed to optimize the base learners in tiers 1 and 3 for model performance enhancement. A comprehensive and robust IDS with both known and unknown attack detection functionalities can be obtained after the model learning and optimization procedures. Additionally, the quality of the used datasets can be improved by data pre-processing and feature engineering procedures to achieve more accurate attack detection.

The performance of the proposed MTH-IDS is evaluated on two public network datasets, the CAN-intrusion-dataset \cite{GIDS} and the CICIDS2017 dataset \cite{CIC}, representing the intra-vehicle and external network traffic data, respectively. The model's feasibility, effectiveness, and efficiency are evaluated using various metrics, including accuracy, detection rates, false alarm rates, F1-scores, and model execution time.

To our knowledge, no previous work proposed such a hybrid IDS that optimizes learning models to accurately detect existing and zero-day attack patterns on both intra-vehicle and external vehicular networks.

The main contributions of this paper are as follows:
\begin{enumerate}
\item It proposes a novel multi-tiered hybrid IDS that can accurately detect the various surveyed types of cyber-attacks launched on both intra-vehicle and external vehicular networks;
\item It proposes a novel feature engineering model based on information gain (IG), fast correlation-based filter (FCBF), and kernel principal component analysis (KPCA) algorithms; 
\item It proposes a novel anomaly-based IDS based on CL-k-means and biased classifiers to detect zero-day attacks;
\item It discusses the use of Bayesian optimization techniques to automatically tune the parameters of each tier in the proposed IDS for model optimization;
\item It evaluates the performance and overall efficiency of the proposed model on two state-of-the-art datasets, CAN-intrusion-dataset and CICIDS2017, and discusses its feasibility in real-world IoV devices. 
\end{enumerate}

The remainder of this paper is organized as follows: Section II discusses the related work. Section III presents the vulnerabilities of intra-vehicle and external vehicular networks, as well as the attack scenarios and IDS deployment.
In Section IV, all the tiers and algorithms in the proposed MTH-IDS are discussed in detail. Section V presents and discusses the experimental results. Section VI concludes the paper.

\section{Related Work}
\subsection{CAN Intrusion Detection}
The research on IDS development for IoV and connected vehicles has been considered critical in recent years. Many research works have a focus on detecting attacks on CAN-based intra-vehicle networks. 
Alshammari \textit{et al.} \cite{known1} proposed an intrusion classification model to identify CAN intrusions on in-vehicle networks utilizing support vector machine (SVM) and k-nearest neighbors (KNN) algorithms. 
Barletta \textit{et al.} \cite{can-mdpi} proposed a distance-based IDS for CAN intrusion detection using a X–Y fused Kohonen network with the k-means algorithm (XYF-K). The proposed method shows high accuracy on the CAN-intrusion dataset, but its main limitation is the high computational complexity.  
Olufowobi \textit{et al.} \cite{reco1} proposed a specification-based real-time IDS named SAIDuCANT to detect the in-vehicle network attacks. The effectiveness of SAIDuCANT is evaluated on a synthetic dataset and the CAN-intrusion dataset. 
Olufowobi \textit{et al.} \cite{reco2} proposed an anomaly-based IDS for CAN attack detection using the adaptive cumulative sum (CUSUM) algorithm. This technique can effectively detect intrusions with low delay based on statistical changes.
Lee \textit{et al.} \cite{otids} proposed an Offset Ratio and Time Interval based IDS (OTIDS) to detect CAN attacks in in-vehicle networks. They also created a CAN dataset by simulating DoS, fuzzy, and impersonation attacks for IDS evaluation.

Deep learning (DL) methods are also widely used for intra-vehicle network IDS development. 
Lokman \textit{et al.} \cite{known3} proposed an unsupervised DL-based anomaly detection model named stacked sparse autoencoders (SSAEs) to discover anomalies in CAN-bus data for intra-vehicle network security enhancement. 
Song \textit{et al.} \cite{known4} proposed a deep convolutional neural network (DCNN) method named Reduced Inception-ResNet to detect intra-vehicle attacks and achieve high detection performance on the CAN-intrusion dataset.   
Ashraf \textit{et al.} \cite{tits} proposed a DL-based IDS for IoV using a long-short term memory (LSTM) autoencoder algorithm. The effectiveness of the proposed model is evaluated on the CAN-intrusion-dataset and UNSW-NB15 dataset, representing in-vehicle network and external network datasets, respectively. DL methods can often achieve high accuracy, but they are computationally expensive due to high model complexity.

\subsection{External Network Intrusion Detection}
Intrusion detection in IoV or external vehicular networks has also attracted significant attention. 
Alheeti \textit{et al.} \cite{defense} proposed an intelligent IDS using back-propagation neural networks to detect DoS attacks in external vehicular networks using the Kyoto 2006+ dataset, but did not consider other attack types. 
Rosay \textit{et al.} \cite{iovcic1} proposed a multi-layer perceptron (MLP) based network IDS for cyber-attack detection in IoT and connected vehicles. The proposed model has been implemented on an automotive microprocessor, and its performance is evaluated on the two variants of the CICIDS2017 dataset. 
Aswal \textit{et al.} \cite{cic_bot} analyzed the applicability of six classical ML algorithms for Bot attack detection on IoV. They used the Bot attack files in the CICIDS2017 dataset to represent the Botnets in vehicular networks, but they did not consider other attacks. 
Aloqaily \textit{et al.} \cite{iov_d1} proposed a network IDS for IoV and connected vehicles using deep belief network (DBN) and decision tree (DT) algorithms. This method shows high accuracy on the NSL-KDD dataset. 
Gao \textit{et al.} \cite{iov_d2} proposed a distributed network IDS for distributed DoS (DDoS) attack detection in vehicular networks and V2X systems. Two general network benchmark datasets, the NSL-KDD and UNSW-NB15 datasets, are used to present the vehicular network datasets and evaluate the IDS.
Schmidt \textit{et al.} \cite{iov_d3} proposed a spline-based IDS for vehicular networks using the knot flow classification (KFC) method and used the NSL-KDD dataset to represent vehicle networks for model evaluation. 

Several other research works also pay attention to the IDS development of general networks and use benchmark datasets for method evaluation. 
Min \textit{et al.} \cite{known3c} proposed a semi-supervised learning model, named SU-IDS, by combining the auto-encoder algorithm with k-means to detect cyber-attacks using the NSL-KDD and CICIDS2017 datasets. 
Yao \textit{et al.} \cite {known4c} proposed a DL model named spatial-temporal deep learning on communication graphs (STDeepGraph) by combining the convolutional neural network (CNN) and long short-term memory (LSTM) methods. The performance of STDeepGraph is evaluated on the UNSW-NB15 and CICIDS2017 datasets.
Injadat \textit{et al.} \cite{MLIDS} proposed a novel multi-stage optimized ML-based IDS for network attack detection and evaluated the model’s performance on the CICIDS2017 and UNSW-NB15 datasets. The ML models used in this paper are optimized by hyper-parameter optimization (HPO) methods.

\subsection{Literature Comparison}
Although various studies about vehicular network IDS development have been published, most of them are only designed for known attack detection on either intra-vehicle (\cite{known1}-\cite{known4}) or external networks (\cite{defense}-\cite{known3c}). Additionally, several papers \cite{defense} \cite{cic_bot} \cite{iov_d2} only consider a specific type of attack, like Botnets or DoS attacks. However, in real-world applications, both intra-vehicle and external networks are vulnerable to various types of attacks with both existing and new patterns. The IDS proposed in \cite{tits} is the only research that considers both CAN bus and external networks, and the IDS proposed in \cite{defense} is the only technique that can detect both known and unknown attacks. Thus, there still should an IDS designed for the detection of both known and zero-day attacks on both intra-vehicle and external vehicular networks. Our proposed IDS aims to achieve this. 

On the other hand, for the deployment of IDSs in real-world vehicle systems, vehicle-level model testing and real-time analysis should be performed to validate the feasibility of the IDSs. However, only five papers \cite{reco1} \cite{otids} \cite{known4} \cite{iovcic1} \cite{iov_d1} did vehicle-level testing or real-time analysis, so the feasibility of other techniques in real-world IoVs is not proven. Therefore, our proposed IDS has been evaluated in a vehicle-level machine to verify whether it meets the real-time requirements of vehicular networks. 

An effective IDS should achieve a high detection rate and a low false alarm rate. Moreover, to meet the real-time requirements of IoV, an IDS should have low computational complexity and high efficiency. Thus, three important procedures, including data sampling, feature engineering, and model optimization, are implemented in our proposed MTH-IDS to improve the efficiency and accuracy of IoV attack detection. The details of how these three procedures can improve the model’s performance are provided in Sections IV-B to IV-D. However, only six existing techniques \cite{known1} \cite{tits}-\cite{cic_bot} \cite{MLIDS} performed feature engineering and only one research work \cite{MLIDS} implemented model optimization. The use of efficiency and accuracy enhancement techniques enables our proposed MTH-IDS to outperform most existing research works in terms of detection rate, false alarm rate, and execution speed. 

To summarize, the main functionalities and components of the proposed MTH-IDS are listed in Table I. Table I also clearly compares 17 existing literature with the proposed MTH-IDS based on the important functionalities that an effective and efficient vehicular network IDS should have.

\begin{table*}[tbp]
\caption{Comparison of Recent Intrusion Detection Techniques for IoVs}
\centering
\setlength\extrarowheight{1pt}
\scalebox{0.84}{
\begin{tabular}{|>{\centering\arraybackslash}m{9.3em}|>{\centering\arraybackslash}m{7em}|>{\centering\arraybackslash}m{7em}|>{\centering\arraybackslash}m{7.5em}|>{\centering\arraybackslash}m{5.5em}|>{\centering\arraybackslash}m{8.2em}|>{\centering\arraybackslash}m{6em}|>{\centering\arraybackslash}m{5.5em}|>{\centering\arraybackslash}m{5.7em}|}

\hline
\textbf{Paper}                                      & \textbf{In-vehicle Network Attack Detection} & \textbf{External Network Attack Detection} & \textbf{Multiple Types of Known Attack Detection} &\textbf{ Zero-Day Attack Detection} & \textbf{Vehicle-Level Model Testing or Real-time Analysis} & \textbf{Data Pre-Processing and Sampling} & \textbf{Feature Engineering} & \textbf{Model Optimization} \\ \hline
Alshammari   \textit{et al.} \cite{known1} & \Checkmark                             &                                   & \Checkmark                               &                           &                                                   &                                  & \Checkmark          &                    \\ \hline
Barletta   \textit{et al.} \cite{can-mdpi}      & \Checkmark                             &                                   & \Checkmark                               &                           &                                                   &                                  &                     &                    \\ \hline
Olufowobi   \textit{et al.} \cite{reco1}        & \Checkmark                             &                                   & \Checkmark                               &                           & \Checkmark                                        &                                  &                     &                    \\ \hline
Olufowobi   \textit{et al.}\cite{reco2}         & \Checkmark                             &                                   & \Checkmark                               &                           &                                                   &                                  &                     &                    \\ \hline
Lee \textit{et al.} \cite{otids}                & \Checkmark                             &                                   & \Checkmark                               &                           & \Checkmark                                        &                                  &                     &                    \\ \hline
Lokman   \textit{et al.} \cite{known3}     & \Checkmark                             &                                   & \Checkmark                               &                           &                                                   &                                  &                     &                    \\ \hline
Song   \textit{et al.} \cite{known4}       & \Checkmark                             &                                   & \Checkmark                               &                           & \Checkmark                                        &                                  &                     &                    \\ \hline
Ashraf \textit{et al.} \cite{tits}            & \Checkmark                             & \Checkmark                        & \Checkmark                               &                           &                                                   &                                  & \Checkmark          &                    \\ \hline
Alheeti   \textit{et al.} \cite{defense}   &                                        & \Checkmark                        &                                          & \Checkmark                &                                                   &                                  & \Checkmark          &                    \\ \hline
Rosay \textit{et al.} \cite{iovcic1}          &                                        & \Checkmark                        & \Checkmark                               &                           & \Checkmark                                        &                                  & \Checkmark          &                    \\ \hline
Aswal \textit{et al.} \cite{cic_bot}          &                                        & \Checkmark                        &                                          &                           &                                                   &                                  & \Checkmark          &                    \\ \hline
Aloqaily \textit{et al.} \cite{iov_d1}          &                                        & \Checkmark                        & \Checkmark                               &                           & \Checkmark                                        &                                  &                     &                    \\ \hline
Gao \textit{et al.}   \cite{ iov_d2}           &                                        & \Checkmark                        &                                          &                           &                                                   &                                  &                     &                    \\ \hline
Schmidt   \textit{et al.} \cite{iov_d3}        &                                        & \Checkmark                        & \Checkmark                               &                           &                                                   &                                  &                     &                    \\ \hline
Min   \textit{et al.} \cite {known3c}      &                                        & \Checkmark                        & \Checkmark                               &                           &                                                   &                                  &                     &                    \\ \hline
Yao   \textit{et al.} \cite {known4c}      &                                        & \Checkmark                        & \Checkmark                               &                           &                                                   &                                  &                     &                    \\ \hline
Injadat \textit{et al.} \cite{MLIDS}         &                                        & \Checkmark                        & \Checkmark                               &                           &                                                   &                        & \Checkmark          & \Checkmark         \\ \hline
Proposed MTH-IDS                           & \Checkmark                             & \Checkmark                        & \Checkmark                               & \Checkmark                & \Checkmark                                        & \Checkmark                       & \Checkmark          & \Checkmark         \\ \hline
\end{tabular}
}
\end{table*}

\section{Vehicular Networks, Vulnerabilities, and IDS Deployment}
\subsection{Vulnerabilities of Intra-vehicle Networks}
Modern vehicles often contain 70-100 ECUs that are in-vehicle components used to enable various functionalities \cite{ECU}. CAN \cite{GIDS} is a bus communication protocol that defines an international standard for efficient and reliable intra-vehicle communications among ECUs. A CAN-bus is built based on differential signaling and comprises a pair of channels, CAN-High and CAN-Low, representing the two signals, 1 and 0, respectively \cite{CANsignal}. CAN is the most common type of IVN due to its low cost and complexity, high reliability, noise-resistance, and fault-tolerance properties \cite{globecom} \cite{GIDS}. However, CAN is vulnerable to various cyber threats due to its broadcast transmission strategy, lack of authentication and encryption, and unsecured priority scheme \cite{CANrisk}.

\begin{figure}
     \centering
     \includegraphics[width=8.5cm]{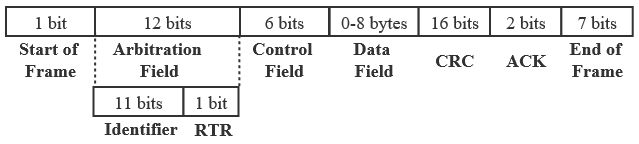}
     \caption{CAN packet structure.} \label{packet}
\end{figure}

CAN messages, or packets, are transmitted via CAN-bus. The data frame is the most important type of CAN packet used to transmit user data \cite{CANframe}. Fig. \ref{packet} shows the structure of a CAN packet, which consists of seven fields \cite{V2XCAN}: start of frame, arbitration field, control field, data field, CRC (cyclic redundancy code) field, acknowledge (ACK) field, and end of frame. Among all fields, the data field with the size of 0-8 bytes is the most important and vulnerable one, since it contains the actual transmitted data that determines the node actions \cite{GIDS}. An attacker can intrude or take control of a vehicle by injecting malicious messages into the data field of CAN packets, resulting in compromised nodes or vehicles; so-called message injection attacks. 

Message injection attacks are the primary type of intra-vehicle attack and can be further classified as DoS attacks, fuzzy attacks, and spoofing attacks by their objectives \cite{GIDS}. In DoS attacks, a CAN is flooded with massive high-priority messages to cause latencies or unavailability of other legitimate messages. Similarly, fuzzy attacks can be launched by injecting arbitrary messages with randomly spoofed identifiers or packets, causing compromised vehicles to exhibit unintended behaviors, like sudden braking or gear shift changes. Spoofing or impersonation attacks, such as gear spoofing and revolutions per minute (RPM) spoofing attacks, are launched by injecting messages with certain CAN identifiers (IDs) to masquerade as legitimate users and take control of the vehicles. 

\subsection{Vulnerabilities of External Vehicular Networks }
In a similar fashion, V2X technology enables interactions and communications between vehicles and other IoV entities, including pedestrians, infrastructures, smart devices, and network systems \cite{V2XCAN} \cite{fog}. With the increasing connectivity of modern IoV, external vehicular networks are becoming large networks that involve various other networks and devices. Thus, external vehicular networks are vulnerable to various general cyber threats because each vehicle or device is a potential entry point for intrusions. Typical attacks in IoV include DoS, GPS spoofing, jamming, sniffing, brute-force, Botnets, infiltration, and web attacks \cite{CIC} \cite{attack1}. The description and IoV scenarios of these common external vehicular network attacks are summarized in Table \ref{external}.
\begin{table}[]
\caption{Common Attack Types on External Vehicular Networks}
\centering
\setlength\extrarowheight{1pt}

\scalebox{0.85}{
\begin{tabular}{|>{\centering\arraybackslash}m{6.5em}|m{27em}|}

\hline
\textbf{Attack Type}           & {\centering\arraybackslash}{\textbf{Description and IoV Scenarios}}                                                                                                                    \\ \cline{1-2}
{DoS \cite{attack1}}         & Send a large number of requests to exhaust the compromised nodes' resources, causing vehicle unavailability or accidents.                                                    \\ \cline{1-2}
GPS Spoofing \cite{attack1} & Masquerade as authorized IoV users to provide a node with false information, like false geographic information, therefore causing fake evidence, event delay, or property losses. \\ \cline{1-2}
Jamming \cite{attack1}      & Jam signals to prevent legitimate IoV devices from communicating with connected vehicles.                                                                    \\ \cline{1-2}
{Sniffing \cite{attack1}}     & Capture vehicular network packets to steal confidential or sensitive information of vehicles, users, or enterprises.                                                      \\ \cline{1-2}
Brute-force \cite{CIC}  & Crack passwords in vehicle systems to take control of vehicles or machines and perform malicious actions.                                                             \\ \cline{1-2}
{Botnets \cite{CIC}}      & Infect multiple connected vehicles and IoV devices with Bot viruses to breach them and launch other attacks.                                                                     \\ \cline{1-2}
Infiltration \cite{CIC} & Traverse the compromised vehicle systems and create a backdoor for future attacks.                                                                            \\ \cline{1-2}
Web Attack \cite{CIC}  & Hack IoV servers or web interfaces of connected vehicles to gain confidential information or perform malicious actions.                                                               \\ \cline{1-2}
\end{tabular}
}
\label{external}%
\end{table}

\begin{figure}
     \centering
     \includegraphics[width=9cm]{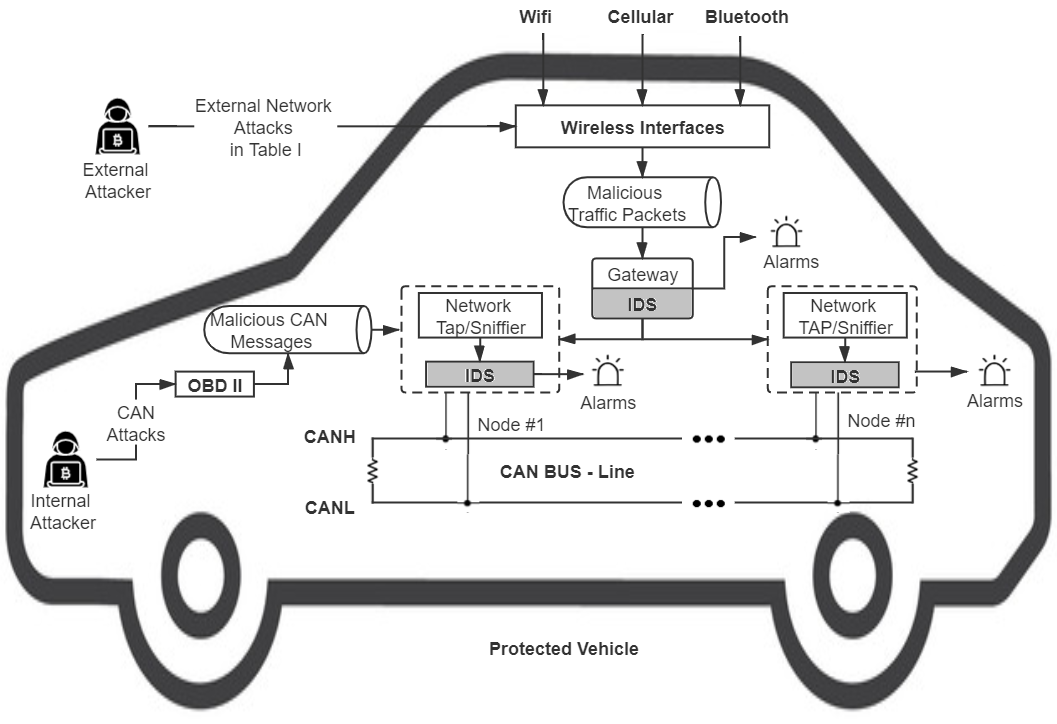}
     \caption{The proposed IDS-protected vehicle architecture.} \label{im}
\end{figure}

\subsection{Attack Scenarios and IDS Deployment}
The attack scenarios and general architecture of a vehicle protected by IDS are shown in Fig. \ref{im}. As the physical interface for ECU communications, the on-board diagnostics II (OBD-II) interface is often exploited by internal attacks to inject malicious CAN messages into the vehicle systems, therefore taking control of the CAN nodes to perform malicious actions, like sudden braking. On the other hand, the external attackers can launch the cyber-attacks listed in Table \ref{external} through various wireless interfaces, including WiFi, cellular network, and Bluetooth \cite{tits}. 

To secure IoV, the proposed IDS can be placed in different locations in vehicular networks. In intra-vehicle networks, the IDS can be deployed on top of the CAN-bus to detect malicious messages \cite{install1}. Since every message is broadcasted to all nodes, it will also be transmitted through the IDS when its signal changes from CAN-High to CAN-Low. The IDS will monitor the CAN packets and identify potential intrusions. If an attack is detected, alarms will be triggered on every node.

On the other hand, the central gateways are common network devices to incorporate IDSs, so the proposed IDS can be placed inside the gateway to monitor the external network traffic \cite{install2}. Thus, abnormal network traffic can be detected when it is passed through the gateways in external vehicular networks. The potential deployment of the proposed IDS is also shown in Fig. \ref{im}.

To protect the vehicle from being breached, all packets or messages transmitted in the protected vehicular network are captured by packet taps or sniffers (\textit{e.g.}, NetFlow), and then analyzed by the proposed IDS before being forwarded to the protected vehicle \cite{sniffer}. For example, if an attacker is launching a DoS attack by sending a large volume of malicious traffic to a vehicle system, the malicious traffic will be detected by the proposed IDS by processing the traffic data captured by the sniffer; alarms will then be generated, and the attacker’s access will be denied, therefore protecting the vehicle from being breached \cite{SDP1}-\cite{V2XAM2}. 

\subsection{Real-Time Requirements of Vehicle Systems}
The standardized vehicular communications specify the performance requirements of vehicle safety services based on two primary metrics: packet-delivery-ratio (PDR) and latency \cite{realrequire}. For IDS development, the latency metric should be considered to meet the real-time requirements. Latency indicates the time needed to transmit a packet from its source to its destination. According to the United States (US) department of transportation, the highest priority vehicle safety services, like collision and attack warnings, should have a latency of 10 to 100 ms at the utmost \cite{realrequire}. On the other hand, for autonomous or cooperative driving, the V2X traffic safety applications require a stringent latency requirement of 10 to 20 ms \cite{V2XAM2} \cite{V2XAM1}. Thus, for a vehicle-level IDS, the time needed to process each network packet is required to be less than 10ms to meet the real-time or latency requirements. 

\section{Proposed MTH-IDS Framework}
\subsection{System Architecture}
\begin{figure*}
     \centering
     \includegraphics[width=14cm]{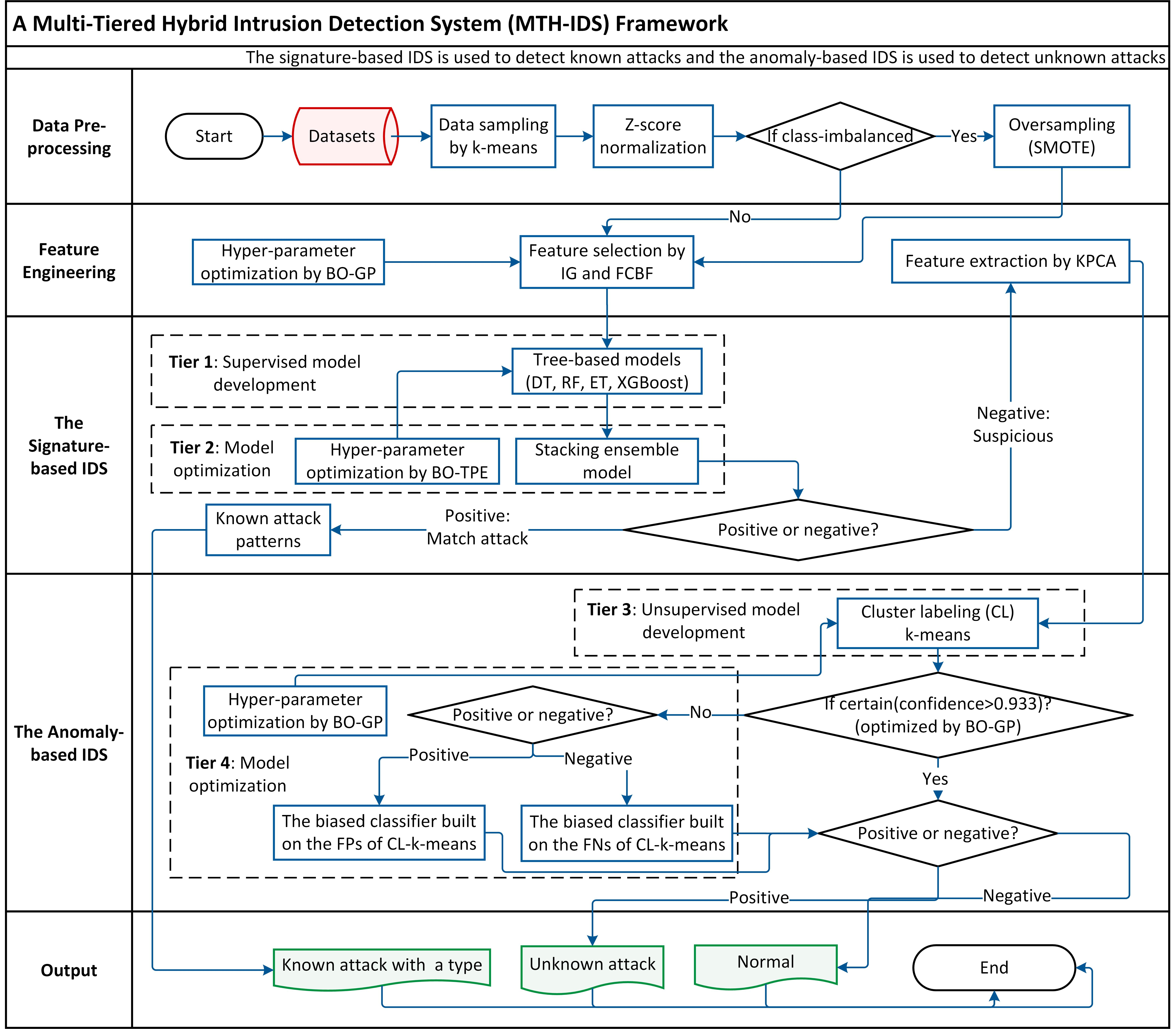}
     \caption{The framework of the proposed MTH-IDS. } \label{frame}
\end{figure*}

The purpose of this work is to develop an IDS that can protect both intra-vehicle and external networks from being breached by the various common attacks presented in Sections III-A and III-B. In this paper, a novel multi-tiered hybrid IDS is proposed to detect both known and unknown cyber-attacks on vehicular networks with optimal performance. Fig. \ref{frame} demonstrates the architecture of the proposed system, comprising four main stages: data pre-processing, feature engineering, a signature-based IDS, and an anomaly-based IDS. 

Firstly, intra-vehicle and external network traffic datasets are collected for the purpose of system performance evaluation on both types of vehicular networks. Data pre-processing consists of a k-means-based cluster sampling method used to generate a highly-representative subset, and a SMOTE method used to avoid class-imbalance. In the feature engineering process, the datasets are processed by information-gain-based and correlation-based feature selection methods to remove irrelevant and redundant features, and then passed to the KPCA model to further reduce dimensionality and noisy features. The proposed data pre-processing and feature engineering procedures can greatly improve the quality of the network data for more accurate model learning. The signature-based IDS is then developed to detect known attacks by training four tree-based machine learners as the first tier of the proposed MTH-IDS: DT, RF, ET, XGBoost. In the second tier, a stacking ensemble model and the BO-TPE method are used to further improve the intrusion detection accuracy by combining the output of the four base learners from the first tier and optimizing the learners. In the next stage, an anomaly-based IDS is constructed to detect unknown attacks. In the anomaly-based IDS, the suspicious instances are passed to a cluster-labeling (CL) k-means model as the third tier to effectively separate attack samples from normal samples. The fourth tier of the MTH-IDS comprises the BO-GP method and two biased classifiers used to optimize the model and reduce the classification errors of the CL-k-means. 
Ultimately, the detection result of each test sample is returned, which could be a known attack with its type, an unknown attack, or a normal packet. 
To summarize the rationale behind the algorithms used in the proposed IDS, the brief description and performance impact of each algorithm are presented in Table \ref{impact}. A detailed description is provided in Sections IV-B to IV-D.

\begin{table*}
\centering
\caption{Rationale and Performance Impact of Each Component of the MTH-IDS}%
\setlength\extrarowheight{1pt}
\scalebox{0.85}{
\begin{tabular}{|>{\centering\arraybackslash}m{6em}|>{\centering\arraybackslash}m{6em}|>{\arraybackslash}m{37em}|>{\arraybackslash}m{20em}|}
\hline
\textbf{Stage}                                & \textbf{Algorithm}                                      & \textbf{Rationale and Description}                                                                                     & \textbf{Performance Impact}                                    \\ 
\hline
\multirow{3}{7em}{Data pre-processing}     & K-means cluster sampling                       & Network traffic data is often large, while IoV devices often have limited computational power and resources. The k-means sampling method can generate highly representative subsets for more efficient training because the removed data is mostly redundant data.                                              & Improve model training efficiency.                     \\ 
\cline{2-4}
                                         & SMOTE                                          & Network traffic data is often imbalanced data because most data samples are collected under normal conditions in real-world vehicle systems. SMOTE can create high-quality samples for minority classes to avoid class-imbalance and ineffective classifiers.                       & Improve detection rate.                                \\ 
\cline{2-4}
                                         & Z-score                         & Different features often have different ranges, which can bias the model training. The Z-score method can normalize features to a similar scale and handle outliers.         & Improve model accuracy and training
  efficiency.      \\ 
\hline
\multirow{3}{7em}{Feature engineering}     & IG                                             & For certain tasks like intrusion detection, many collected features can be irrelevant, causing additional training time. The IG method can remove those unimportant features.                                                      & Improve model training
  efficiency.                   \\ 
\cline{2-4}
                                         & FCBF                                           & Certain features are redundant because they contain very similar information. FCBF can remove redundant features by calculating the correlation between each pair of features.                                                             & Improve model accuracy and training
  efficiency.      \\ 
\cline{2-4}
                                         & KPCA                                            & The anomaly-based IDSs are sensitive to the quality of features. KPCA can further extract the most relevant features to reduce dimensionality and noisy information.                                                 & Improve model accuracy and training
  efficiency.      \\ 
\hline
\multirow{3}{7em}{The signature- based IDS} & DT, RF,
  ET, and XGBoost & Tree-based ML algorithms often perform better than other ML algorithms on complex tabular data to which IoV data belong. Four tree-based supervised algorithms are used to train base classifiers for known intrusion detection.                              & Detect various types of known attacks.                                  \\ 
\cline{2-4}
                                         & BO-TPE                                         & The default hyper-parameters of ML algorithms often cannot return the best model. BO-TPE can optimize the models' hyper-parameters to obtain the optimized base classifiers.                                                              & Improve accuracy of known
  attack detection.    \\ 
\cline{2-4}
                                         & Stacking                                       & Ensemble models can often achieve higher accuracy than any single model. Stacking ensemble can combine the base classifiers to obtain a meta-learner with better performance.                                           & Improve accuracy of known
  attack detection.    \\ 
\hline
\multirow{3}{7em}{The anomaly- based IDS}   & CL-k-means                                     & For unknown attack detection, CL-k-means can generate a sufficient number of normal and attack clusters to identify zero-day attacks from the newly arriving data.                 & Detect unknown attacks.                                \\ 
\cline{2-4}
                                         & BO-GP                                          & CL-k-means has an important hyper-parameter, the number of clusters, $k$. BO-GP is an effective HPO method to optimize $k$ and obtain the optimized CL-k-means
  model.                                                           & Improve accuracy of
  unknown attack detection.  \\ 
\cline{2-4}
                                         & Two biased classifiers                         & CL-k-means may return many errors when detecting complex unknown attacks. Two biased classifiers are trained on the FPs and FNs of CL-k-means to reduce the errors. & Improve accuracy of unknown attack detection.  \\
\hline
\end{tabular}
}
\label{impact}%
\end{table*}

\subsection{Data Pre-processing}
\subsubsection{Data Sampling by K-means Clustering}
In real life, training ML models on massive amounts of network traffic data is unrealistic and may cost a massive amount of time, especially in the hyper-parameter tuning process that needs to train a ML model multiple times. For model training efficiency improvement purposes, data sampling is a common technique that can generate a subset of the original data to reduce the training complexity of a model \cite{cluster}. 

In the proposed system, to obtain a highly-representative subset, a k-means-based cluster sampling method is utilized. Cluster sampling is a common data sampling method by which the original data points are grouped into multiple clusters; then, a proportion of data is sampled from each cluster to form a representative subset \cite{cluster}. Unlike random sampling, which randomly selects every data sample with an equal probability, cluster sampling can generate a highly-representative subset because the discarded data points are mostly redundant data. 

Among all clustering algorithms, k-means is the most common one for data sampling due to its simple implementation and low computational complexity \cite{kmeq}. K-means clustering algorithms are used to divide the data points into $k$ clusters based on their Euclidean, Manhattan, or Mahalanobis distances \cite{kmour} \cite{AM_super}. The data samples in the same group can be considered similar samples, so sampling from each group can greatly reduce the size of data without losing important information. K-means aims to minimize the sum of squares of distances between all the data points and the corresponding centroid of the cluster, denoted by \cite{kmeq}:
\begin{equation}
\sum_{i=0}^{n_{k}} \min _{u_{j} \in C_{k}}\left(\mathbf{x}_{i}-u_{j}\right)^{2},\end{equation}
where \begin{math}(\mathbf{x}_1,\cdots, \end{math}\begin{math} \mathbf{x}_n)\end{math} is the data matrix; $u_j$, also called the centroid of a cluster $C_k$, is the mean of all the samples in $C_k$; and $n_k$ is the total number of sample points in the cluster $C_k$. K-means has a linear time complexity of $O(nkt)$, where $n$ is the data size, $k$ is the number of clusters, and $t$ is the number of iterations \cite{kmeq}.

After implementing the k-means to cluster the original data samples into $k$ clusters, random sampling is then applied to each cluster to select 10\% of the data as the sampled subset. The percentage of data sampling can vary, depending on the data scale and resource limitations. Additionally, the main hyper-parameter of k-means, the number of clusters \cite{hpme}, $k$, is tuned by Bayesian Optimization (BO) to improve the quality of the subset. 

Hyper-parameter optimization (HPO) is the process of building an optimized ML model for a specific problem or dataset using an optimization algorithm \cite{IDP2AM}. BO algorithms are an efficient group of HPO algorithms that determine the next hyper-parameter value based on the previous evaluation results \cite{AMBO}. In BO, a surrogate model is used to fit all the currently tested data points into the objective function; an acquisition function is then used to locate the next point. Two common surrogate models for BO are the Gaussian process (GP) and the tree Parzen Estimator (TPE) \cite{hpme}. 

In GP surrogate models, the predictions follow a Gaussian distribution \cite{hpme}: 
\begin{equation}
p(y | x, D)=N\left(y | \hat{\mu}, \hat{\sigma}^{2}\right),
\end{equation}
where $D$ is the hyper-parameter configuration space, $y=f(x)$ is the value of the objective function for each hyper-parameter configuration with its mean as $\mu$ and covariance as $\sigma$. The BO method with GP (BO-GP) exhibits great performance on optimizing a small number of continuous or discrete hyper-parameters due to its fast convergence speed, but is inefficient for conditional hyper-parameters since it treats each hyper-parameter independently \cite{hpme}. BO-GP has a time complexity of $O(n^3)$ and space complexity of $O(n^2)$ \cite{hpme}.

Since k-means methods generally only have one discrete hyper-parameter, $k$, that needs to be tuned, BO-GP serves as an effective HPO method for k-means because of its fast convergence speed. The silhouette coefficient, a common distance-based metric that can effectively evaluate clustering performance, is chosen to be the metric of the k-means model and used as the objective function of BO-GP. It measures how similar a data point is to other data points within the same cluster and how different the data point is from the data points in other clusters \cite{kmour}. 
\subsubsection{Reduce Class-Imbalance by Oversampling}
Class-imbalance issues often occur in network traffic data, since the percentage of normal samples is often much larger than the percentage of attack samples in real-world network data, resulting in biased models and low detection rate \cite{imbalance}. 

Class-imbalance problems are mainly solved by resampling methods, including random sampling and synthetic minority oversampling techniques (SMOTE), which can create new instances for the minority classes to balance the dataset \cite{imbalance}.
Unlike random sampling, which simply replicates the instances and may cause over-fitting, SMOTE \cite{SMOTE} can synthesize high-quality instances based on the concept of KNN; thus, SMOTE is chosen in the proposed IDS to solve class-imbalance. For each instance $X$ in the minority class, assuming $X_i$ is a sample randomly selected from the $k$ nearest neighbors of $X$, a new synthetic instance $X_{n}$ can be denoted by \cite{MLIDS}:
\begin{equation}
X_{n}=X+rand(0,1) *\left(X_{i}-X\right), i=1,2, \cdots, k,
\end{equation}
where $rand(0,1)$ represents a random number in the range of $(0, 1)$.
Thus, SMOTE is utilized in the proposed IDS to create high-quality instances for minority classes. 

\subsubsection{Data Normalization}
After implementing the k-means and SMOTE methods to obtain a representative and balanced dataset, several additional data pre-processing steps are completed for the next steps.  Firstly, the network traffic datasets are encoded with a label encoder used to transform categorical features into numerical features to support the inputs of ML algorithms, because many ML algorithms cannot support string features directly. After that, the network datasets are normalized by the Z-score algorithm since the collected features in network traffic data often have largely different ranges, and ML models often perform better on normalized datasets \cite{defense}. An unnormalized dataset with largely different feature scales may result in a biased ML model that only lays emphasis on large-scale features. Through the Z-score method, the features can be normalized to have a mean of 0 and a standard deviation of 1. By implementing the Z-score method, each normalized feature value, $x_n$, is denoted by:
\begin{equation}
x_{n}=\frac{x-\mu}{\sigma},
\end{equation}
where $x$ is the original feature value, $\mu$ and $\sigma$ are the mean and standard deviation of the feature values, respectively.
\subsection{Feature Engineering}
A high-quality and highly representative dataset can be generated after data pre-processing. On the other hand, obtaining an optimal feature list by appropriate feature engineering can also improve the quality of datasets for more accurate and efficient model learning.
A comprehensive feature engineering method that consists of IG, FCBF, and KPCA,  is implemented before ML model training to remove irrelevant, redundant, and noisy features while retaining the important features \cite{PCAreason}.
\subsubsection{Feature Selection by Information Gain}
As a common feature selection (FS) method, the information gain (IG) method is used to select important features. IG, the amount of information gained or the changes in entropy, can be used to measure how much information a feature can bring to the targeted variable \cite{IG}. IG is chosen in the proposed system since it can obtain an importance score for each feature at a fast speed due to its low computational complexity of $O(n)$ \cite{IG}. The importance score of each feature enables us to select the most relevant features for the task. Assuming $T$ is the target variable, for each feature denoted by a random variable $X$, the IG value of using the feature $X$, is denoted by \cite{IG}:
\begin{equation}
IG(T | X)=H(T)-H(T | X)\label{IGeq},
\end{equation}
where $H(T)$ is the entropy of the target variable $T$, and $H(T|X)$ is the conditional entropy of $T$ over $X$. 

Therefore, the feature importance of a feature $X$, or the correlation between $X$ and the target $T$, can be represented by the IG value of $T$ over $X$, $IG(T|X)$. A feature $X$ is considered more important to the target $T$ than another feature $Y$ if $IG(T|X) > IG(T|Y)$ \cite{IG}. 

To implement the IG-based FS method, the importance of each feature is calculated based on eq. (\ref{IGeq}) and normalized to have a sum of 1.0, denoting the relative importance. The features are then ranked by their importance and are selected from top to bottom until the total importance of selected features reaches the correlation threshold, $\alpha$. The remaining features, with the total feature importance less than $1-\alpha$, are discarded. To obtain an appropriate correlation threshold, $\alpha$ is optimized by BO-GP that uses the validation accuracy as the objective function for HPO. After this process, the irrelevant features are eliminated, and a reduced number of informative features with high importance are obtained for the next step. 
\subsubsection{Fast Correlation Based Filter (FCBF)}
Although the IG-based FS method eliminates the unimportant features to reduce time complexity, many redundant features still exist. Feature redundancy may increase time and space complexity, and degrade model performance by increasing the probability of being misled by noisy data, as well as increasing the risk of over-fitting \cite{PCAreason}. Thus, removing redundant features by calculating the correlations of input features is beneficial for model performance and efficiency. 

Among the correlation-based FS algorithms, the fast correlation-based filter (FCBF) \cite{FCBF} algorithm is selected since it has shown great performance on high dimensional datasets by effectively removing redundant features while retaining informative features, and has a low time complexity of $O(nlogn)$ \cite{FCBFnew}. 
In FCBF, the symmetrical uncertainty (SU) is calculated to measure the correlations between features by normalizing the IG values \cite{FCBF}:
\begin{equation}
SU(X, Y)=2\left[\frac{\mathrm{IG}(X | Y)}{H(X)+H(Y)}\right].
\end{equation}

$SU(X,Y)$ is in the range $[0, 1]$ with the value 1 indicating a perfect correlation between the two features $X$ and $Y$, while the value 0 indicates the two features are fully independent. 

The FCBF method searches the features in the feature space based on their $SU$ values until the entire feature space has been explored. The highly correlated features are regarded as redundant features, and only one of them will be retained. 

In the proposed FS approach, the $SU$ value of each pair of features is calculated as their correlations. The correlation threshold, $\alpha$, is also optimized by BO-GP. When the correlation value between two features is larger than $\alpha$, the one with higher feature importance is retained while the other is discarded. The correlation calculation and feature deletion procedures are repeated until each pair of features in the feature list are not highly correlated ($SU <= \alpha$). The FS model that combines the IG method and the FCBF algorithm is named IG-FCBF.
\subsubsection{Kernel Principal Component Analysis (KPCA)}
Although utilizing IG-FCBF can return a better feature set than only using IG, FCBF has a major limitation that it only calculates the correlation between pairs of features, but does not consider correlations among three or more different features, resulting in undiscovered noisy features \cite{FCBFnew}. On the other hand, the unsupervised learning models in the anomaly-based IDS are more sensitive to appropriate features than supervised learning models since they rely on the changes of feature values instead of the ground truth labels to process data \cite{kmour} \cite{PCAreason}. 
Hence, kernel principal component analysis (KPCA) is utilized after implementing the IG-FCBF method for the anomaly-based IDS. 

PCA \cite{PCA} is a feature extraction algorithm that uses orthogonal transformations to transform a set of correlated features onto a smaller subset of uncorrelated features, named principal components. KPCA is an improved version of PCA that uses the kernel trick to learn a non-linear function or decision boundary to reduce the dimensionality of non-linear data \cite{KPCA}. 
KPCA is selected due to its adaptability to non-linear data, as well as its capacity to reduce computational complexity, the risk of over-fitting, and distracting noise \cite{PCAreason}.

Additionally, the two essential hyper-parameters in KPCA, the number of extracted features and the kernel type, are optimized by the BO-GP method using validation accuracy as the objective function to improve the model performance. It is efficient for BO-GP to optimize these discrete and categorical hyper-parameters. KPCA is used with IG-FCBF to construct the IG-FCBF-KPCA method to obtain an optimal dataset with extracted features as the input of the anomaly-based IDS.
\subsection{The Proposed Hybrid IDS}
IDSs are mainly classified as signature-based IDSs and anomaly-based IDSs. Signature-based IDSs are designed to detect the known attack patterns by training supervised ML models on labeled datasets. However, they often lack the capacity to detect new attack patterns that are not previously stored in the databases \cite{sigIDS}. On the other hand, anomaly-based IDSs can distinguish unknown attack data from normal data by unsupervised learning algorithms based on the assumption that new attack data are more statistically similar to the known attack data than normal data, but they often return many false alarms \cite{assumption}. Thus, a hybrid IDS that consists of a signature-based IDS and an anomaly-based IDS is proposed in this paper to effectively detect both known and zero-day attacks. 
\subsubsection{The Signature-based IDS}
After the data pre-processing and feature engineering procedures, the obtained labeled datasets are trained by an ensemble learning model to develop a signature-based IDS. In the proposed signature-based IDS, four tree-based ML algorithms — decision tree (DT), random forest (RF), extra trees (ET), and extreme gradient boosting (XGBoost) — are selected as the base learners. 

DT \cite{DT} is a common ML algorithm that uses a tree-structure to fit data and make predictions. DT algorithms have multiple hyper-parameters that require tuning, including the tree depth, minimum sample split, minimum sample leaf, maximum sample nodes, and minimum weight fraction leaf, etc. \cite{hpme}. 
RF \cite{RF} is an ensemble learning model that uses the majority voting rule to combine multiple decision tree classifiers, while ET \cite{ET} combines a collection of randomized decision trees built on different subsets of a dataset. XGBoost \cite{XGBoost} is a gradient-boosted decision tree (GBDT) based algorithm designed for speed and performance improvement. For RF, ET, and XGBoost, the hyper-parameters of DT are also the important hyper-parameters for them because they are all constructed by integrating multiple DTs. Additionally, they have an essential hyper-parameter that needs to be tuned, being the number of base DTs to be built for each model, “n\_estimators”, which has a direct impact on model performance. XGBoost has another hyper-parameter, the learning rate, which determines the convergence speed \cite{hpme}.

Assuming the number of instances is $n$, the number of features is $f$, and the number of DTs in ensemble models is $t$, the time complexity of DT, RF, ET, and XGBoost is $O(n^2 f)$, $O(n^2 \sqrt f t)$, $O(nft)$, and $O(nft)$, respectively \cite{globecom}.

The algorithm choosing reasons are as follows \cite{globecom} \cite{driftmag}:
\begin{enumerate}
\item RF, ET, and XGBoost are all ensemble models that combine multiple DTs and can effectively work on non-linear and complex data to which network traffic data belongs; hence, they often perform better than other ML algorithms, like naïve Bayes (NB) and KNN, which often do not perform well on complex datasets. 
\item They enable parallel execution, which significantly reduces model training time and improves efficiency.
\item They calculate feature importance during the model training process, which is beneficial for feature engineering procedures.
\item The tree-based algorithms have randomness in their construction process, which enables us to build a robust ensemble model that has better generalizability than using other ML algorithms.
\end{enumerate}

After obtaining the four tree-based ML models, they are combined using stacking, an ensemble learning method, to improve model performance because the generalizability of a combination of multiple base learners is usually better than that of a single model \cite{stacking}. Stacking is a standard ensemble learning technique that uses the output labels estimated by four base learners (DT, RF, ET, and XGBoost) as the input features to train a strong meta-learner that makes the final prediction  \cite{stacking}. Using stacking can learn the information from all four base learners to reduce the errors of single learners and obtain a more reliable and robust meta-classifier. In the proposed system, the best-performing one among the four base models is chosen as the algorithm to build the meta-learner because it is most likely to achieve the best performance. 

The important hyper-parameters of the four tree-based ML algorithms are optimized by a HPO method, BO with tree-Parzen estimator (BO-TPE). BO-TPE creates two density functions, $l(x)$ and $g(x)$, to act as the generative models for variables. With a pre-specified threshold $y^*$ to separate the relatively good and poor results, the objective function of TPE is modeled by the Parzen windows \cite{hpme}:
\begin{equation}
p(x | y, D)=\left\{\begin{array}{ll}{l(x),} & {\text { if \ } y<y^{*}} \\ {g(x),} & {\text { if \ } y>y^{*}}\end{array}\right.,
\end{equation}
where $l(x)$ and $g(x)$ indicate the probability of detecting the next hyper-parameter value in the well-performing regions and in the poor-performing regions, respectively.
BO-TPE detects the optimal hyper-parameter values by maximizing the ratio $l(x)/g(x)$. The Parzen estimators are organized in a tree structure, so the specified conditional dependencies of hyper-parameters can be retained. Additionally, BO-TPE can optimize all types of hyper-parameters effectively \cite{hpme}. Therefore, BO-TPE is used to optimize the hyper-parameters of the tree-based ML models that have many hyper-parameters.

\subsubsection{The Anomaly-based IDS}
The proposed signature-based IDS can detect multiple types of known attacks effectively. However, attackers can still carry out zero-day attacks that are not included in the known attack patterns and can be misclassified as normal states. Therefore, the instances labeled “normal” by the signature-based IDS will be considered suspicious instances because some of them can be unknown attack samples. A novel anomaly-based IDS architecture is then developed to identify zero-day attacks by processing the suspicious instances.

After feature engineering, the optimized dataset obtained from the output of the IG-FCBF-KPCA method is used to train the anomaly-based IDS. The first tier of the proposed anomaly-based IDS comprises the cluster labeling (CL) k-means developed by improving the k-means model introduced in Section IV-B1. 
Since there are millions of instances collected under many different situations in network traffic datasets, a sufficient number of clusters should be used to distinguish between normal and attack data. The main procedures of the proposed CL-k-means method are as follows: 
\begin{enumerate}
\item Split the dataset into a sufficient number of clusters using k-means.
\item Label each cluster by the majority label of data samples. In each cluster, the class label to which most of the instances belong, “normal” or “attack”, is assigned to this cluster.
\item Label each sample in the test set as “normal” or “attack” based on the label of the cluster that this instance is classified into by k-means.
\item For each test sample $i$ that is classified into a cluster, calculate the percentage of majority class samples in this cluster as the confidence or clustering probability, $p_i$. 
\item Optimize the number of clusters ($k$) and distance metric as the major hyper-parameters of k-means, by the BO-GP algorithm to obtain the optimal CL-k-means model. The validation accuracy on the test set is used as the objective function for BO-GP.
\end{enumerate}

K-means is selected to distinguish between attack and normal data mainly due to the real-time requirements of vehicle-level systems. K-means is computationally faster than most other clustering algorithms because it has a linear time complexity of $O(nkt)$, where $n$ is the data size, $k$ is the number of clusters, and $t$ is the number of iterations \cite{kmeq}. The model training time is further reduced by using mini-batch k-means, which uses randomly sampled subsets as mini-batches in each training iteration \cite{minikmeans}. Additionally, k-means guarantees convergence and easily adapts to new samples.

To increase the detection rate and reduce the false alarm rate of the CL-k-means method, the second tier of the proposed anomaly-based IDS uses two biased classifiers to reduce the false negatives (FNs) and false positives (FPs), respectively. Biased classifiers are developed by the following procedures:
\begin{enumerate}
\item Collect the FNs and FPs obtained from the training set using the proposed CL-k-means method.
\item Select the best-performing singular supervised learning model in the signature-based IDS (\textit{e.g.}, RF) as the algorithm to construct biased classifiers.
\item Train the first biased classifier, $B_1$, on all the FNs along with the same amount of randomly sampled normal data to build a model that aims to reduce FNs.
\item Train the second biased classifier, $B_2$, on all the FPs along with the same amount of randomly sampled attack data to build a model that aims to reduce FPs.
\end{enumerate}

After implementing the proposed CL-k-means model, each data sample whose clustering probability ($p_i$) is less than a threshold $p_i^*$, is regarded as an uncertain instance. The threshold $p_i^*$ is a continuous variable and has been optimized to be 0.933 by BO-GP that uses the validation accuracy as the objective function. After obtaining the two trained classifiers, the uncertain instances will be passed to $B_1$ (if labeled “normal” by the CL-k-means) or $B_2$ (if labeled “attack” by the CL-k-means) to obtain its final classification result. 

The proposed anomaly-based IDS is constructed under the assumption that new attack patterns are unknown and future incoming data samples are unlabeled; hence, only the FNs and FPs obtained during the training phase are used to build biased classifiers. This enables the proposed IDS to detect new attack patterns without additional procedures that are difficult to perform, like constant data labeling and model updates.

Compared to other unsupervised anomaly detection algorithms like isolation forest (iForest) and one-class SVM (OC-SVM) \cite{unsupervisedids}, the proposed CL-k-means method with biased classifiers has the following advantages to achieving high accuracy and efficiency: 
\begin{enumerate}
\item The proposed CL-k-means model has the capacity of using a sufficient number of clusters to model the data samples with various attack and normal patterns. This makes the proposed method have better generalizability and data pattern modeling capability than other outlier detection methods, like iForest and OC-SVM, which are mostly binary models. 
\item The number of clusters in CL-k-means, $k$, is automatically optimized by BO-GP. This enables the proposed model to automatically fit different datasets and tasks according to the complexity of data patterns. 
\item The main difficulty of unknown attack detection is that unsupervised learning models often return more misclassified samples than supervised learning models. Thus, using the biased classifiers can effectively reduce the FPs and FNs because they can learn the patterns of misclassified data samples that are difficult to be identified by the CL-k-means. 
\item The use of the cluster probability, $p_i$, can greatly improve the model efficiency. This is because the new samples that are very similar to existing attack or normal patterns (with high confidence) can be labeled directly, and only the uncertain samples (with relatively low probability) are passed to the biased classifiers for further identification. 
\item The use of mini-batch k-means can significantly reduce the execution time of the MTH-IDS to meet the real-time requirements of IoV.
\end{enumerate}

However, certain attack patterns are very similar to normal patterns, which makes it difficult to distinguish them. Additionally, the samples collected under certain legitimate network events, like crowd events, may still be misclassified as attack samples because the new patterns they have are largely different from existing normal patterns.  Additionally, although k-means is more suitable for IoV IDS development than other unsupervised learning algorithms due to its low complexity, the data distribution modeling limitation of k-means is another potential issue to be better addressed \cite{kmeq}. If time and budget permits, k-means can be replaced by other clustering algorithms with the same cluster labeling technique based on the specific data shape and distribution to further improve the system performance.
Online learning techniques that can keep updating learning models based on the new attack patterns may also improve the generalizability and accuracy of the IDS, which will be our future work. 
\subsection{Runtime Complexity}
The training process of the proposed MTH-IDS can be done at a server machine with high computational speed, while the testing process should be implemented in vehicle systems.
Developing models with low runtime complexity enables the proposed IDS to reduce the latency of vehicle systems and meet real-time requirements. In the implementations, each test sample will be passed through the stacking model constructed with four tree-based algorithms, the CL-k-means method, and one of the biased classifiers. 
Since the runtime complexity of DT is $O(df)$ and the runtime complexity of RF, ET, XGBoost is $O(dft)$, where $d$ is the maximum depth of the trees, $f$ is the number of features and $t$ is the number of trees, the maximum runtime complexity of the signature-based IDS is only $O(dft)$ \cite{runcom}. For the proposed CL-k-means method in the anomaly-based IDS, its runtime complexity is $O(fk)$, where $k$ is the number of clusters \cite{kmeansrun}. The biased classifier in the anomaly-based IDS is also the best-performing tree-based algorithm, so its maximum runtime complexity is also $O(dft)$. 

Therefore, the maximum overall runtime complexity of the proposed IDS is at a low-level, only $O(2dft+fk)$, since the values of $d$, $f$, $t$, and $k$, are a few dozens at most. The model test time will be calculated in the experiments to evaluate the feasibility of the proposed IDS in vehicle systems. 

\subsection{Validation Metrics}
To evaluate the generalizability of the proposed framework and avoid over-fitting, both cross-validation and hold-out methods are used in the known attack detection experiments. Specifically, the train-test-validation split and model evaluation procedures for each dataset are as follows:
\begin{enumerate}
\item Use 70\%-30\% train-test split to generate a training set with 70\% of data samples and a test set with 30\% of data samples. The test set will remain untouched before the final hold-out validation.
\item Implement 10-fold cross-validation on the training set to evaluate the proposed model on different regions of the dataset. In each iteration/fold of the 10-fold cross-validation, 90\% of the original training set is used for model training, and 10\% of the original training set is used as the validation set for model testing. 
\item Test the trained model obtained from Step 2 on the untouched test set to evaluate the model performance on a new dataset.
\end{enumerate}

70\%-30\% train-test split and 10-fold cross-validation were chosen because they are standard and sufficient numbers to construct powerful validation methods to avoid over-fitting and concept drift issues \cite{split}. If the proposed method can accurately detect intrusions in both cross-validation and hold-out validation, there are no underfitting or overfitting issues \cite{validation}.

On the other hand, hold-out validation is used to evaluate the proposed system for unknown attack detection. For each attack type as a zero-day attack, a validation set consisting of all the instances of this attack type and the same amount of randomly sampled normal data is generated; all the other samples are used as the training set. Therefore, the validation results can be used to evaluate whether the proposed system has the capacity to detect the patterns of each type of unknown attack. This validation process is based on the assumption that new attack data is more statistically similar to certain other known attack data than normal data \cite{assumption}.

Several metrics, including accuracy (Acc), detection rate (DR), false alarm rate (FAR), and F1-score, are used to comprehensively evaluate the performance of the proposed IDS \cite{MLIDS}. By calculating the true positives (TPs), true negatives (TNs), false positives (FPs), and false negatives (FNs) of the proposed model, the used metrics are calculated by the following equations \cite{MLIDS}:
\begin{equation}
Acc= \frac{T P+T N}{T P+T N+F P+F N} 
\end{equation}
\begin{equation}
D R=\frac{T P}{T P+F N} 
\end{equation}
\begin{equation}
F A R=\frac{F P}{T N+F P} 
\end{equation}
\begin{equation}
F 1=\frac{2 \times T P}{2 \times T P+F P+F N} 
\end{equation}

Model execution time, calculated by the average of the model training and validation time in the 10-fold cross-validation or hold-out validation, is used to evaluate model efficiency. An efficient IDS should be able to achieve a high F1-score and low execution time simultaneously.

\section{Performance Evaluation}
\subsection{Experimental Setup}
To develop the proposed MTH-IDS, the feature engineering and ML algorithms were implemented using the Pandas \cite{pandas}, Scikit-learn \cite{sklearn}, Xgboost \cite{xgb} libraries in Python, while the HPO methods were implemented by extending the Skopt \cite{skopt} and Hyperopt \cite{hyperopt} libraries\footnote{
code for the major modules is available at: https://github.com/Western-OC2-Lab/Intrusion-Detection-System-Using-Machine-Learning}. The experiments were carried out on a Dell Precision 3630 Tower machine with an i7-8700 central processing unit (CPU) (6-Core, 3.20 GHz) and 16 Gigabytes (GB) of memory, and a Raspberry Pi 3 machine with a BCM2837B0 64-bit CPU and 1 GB of memory, representing a server machine for model training and a vehicle-level machine for model testing, respectively. 

The experiments are divided into three parts, one for the known intrusion detection by evaluating the signature-based IDS component on the labeled datasets, one for unknown intrusion detection by evaluating the anomaly-based IDS component on the unlabeled datasets, and one for the entire model evaluation by analyzing the CPU resource usage on a vehicle-level machine.

\subsection{Data Description}
For the purpose of intra-vehicle network IDS development, the first used dataset is the CAN-intrusion-dataset proposed in 2018 \cite{GIDS}. The dataset is generated by logging CAN traffic via the OBD-II port of a vehicle when CAN attacks are launched. The features of this dataset include timestamp, CAN ID, data length code (DLC), and the 8-bit data field of CAN packets (DATA[0]-DATA[7]). Since the feature “timestamp” has a strong correlation with cyber-attack simulation periods and can lead to biased models and results, this feature was removed from the feature space. As the authors in \cite{GIDS} created the CAN-intrusion-dataset by injecting attacks into random CAN IDs, this CAN ID feature is retained. Based on the CAN packet structure shown in Fig. \ref{packet}, ten features extracted from the identifier field and data field of CAN packets, including CAN ID, DLC, and DATA[0]-DATA[7], were preliminarily selected for IDS development. The attack types of the CAN-intrusion-dataset are shown in Table \ref{CAN}. Moreover, the label distributions of the original dataset, the training set, and the test set are shown in Table \ref{CAN}. As the minority classes have large numbers of samples (at least 491,847 samples), SMOTE is not required for balancing the CAN-intrusion-dataset.

For external network IDS development, there is a shortage of public IoV benchmark datasets due to popularization, privacy, and commercialization issues \cite{iovcic1} \cite{nodata1} \cite{nodata2}. On the other hand, wireless local area networks (WLANs) and cellular networks are common communication strategies for IoV and connected vehicles, so the attacks launched on conventional computer networks can be considered similar to the intrusions carried out on external vehicular networks \cite{iovcic1} \cite{iovcic2}.  
Therefore, many research projects and papers \cite{tits} - \cite{iov_d3} develop external vehicular network IDSs using general network security datasets, including KDD-99, NSL-KDD, Kyoto 2006+, UNSW-NB15, and CICIDS2017 \cite{iovcic2}. Among these cyber-security datasets, CICIDS2017 \cite{CIC} is the most representative dataset of current external networks because it is the most state-of-the-art dataset and contains more features, instances, and cyber-attack types than other datasets \cite{iovcic1}. Thus, the network traffic flow data in the CICIDS2017 dataset is chosen in the proposed MTH-IDS to represent the complex external vehicular network data. 
Moreover, to better relate the CICIDS2017 dataset to IoV applications, we have associated each type of attack in the CICIDS2017 dataset with the external vehicular network threats described in Table \ref{external} based on the detailed analysis of the CICIDS2017 dataset in \cite{cicdetailed}. The specifics of the CICIDS2017 dataset and the corresponding external vehicular attack types are shown in Table \ref{CIC}. Since the Bot, brute-force, infiltration, and web attack classes are minority classes with small numbers of samples (from 36 to 13,835), the SMOTE method described in Section IV-B2 was implemented to synthesize more samples to enable the minority classes to have at least 100,000 samples. Addressing class-imbalance can avoid obtaining biased models with low attack detection rates.

\begin{table}[tbp] \centering%
\caption{Class Label and Size of The CAN-Intrusion-Dataset}%
\setlength\extrarowheight{1pt}
\scalebox{0.85}{
\begin{tabular}{|>{\centering\arraybackslash}m{7em}|>{\centering\arraybackslash}m{7em}|>{\centering\arraybackslash}m{7.5em}|>{\centering\arraybackslash}m{6.5em}|}
\hline
\textbf{Class Label}   & \textbf{Original Number of Samples} & \textbf{Number of Training Set Samples} & \textbf{Number of Test Set Samples} \\ \hline
Normal          & 14,037,293                    & 9,826,105            & 4,211,188        \\ \hline
DoS             & 587,521                       & 411,265              & 176,256          \\ \hline
Fuzzy           & 491,847                       & 344,293              & 147,554          \\ \hline
RPM   Spoofing  & 654,897                       & 458,428              & 196,469          \\ \hline
Gear   Spoofing & 597,252                       & 418,076              & 179,176          \\ \hline
\end{tabular}%
}
\label{CAN}%
\end{table}%

\begin{table}[tbp] \centering%
\caption{Class Label, Attack Type, and Size of The CICIDS2017 Dataset}%
\setlength\extrarowheight{1pt}
\scalebox{0.85}{
\begin{tabular}{|>{\centering\arraybackslash}m{6em}|>{\centering\arraybackslash}m{6em}|>{\centering\arraybackslash}m{5em}|>{\centering\arraybackslash}m{7em}|>{\centering\arraybackslash}m{4.2em}|}
\hline
\textbf{Class Label }               & \textbf{Corresponding Attack Type in Table \ref{external} \cite{cicdetailed}} & \textbf{Original Number of Samples} & \textbf{Number of Training Set Samples After Balancing} & \textbf{Number of Test Set Samples} \\ \hline
BENIGN                       & -                                             & 2,273,097                     & 1,591,168                               & 681,929                  \\ \hline
Bot                          & Botnets                                       & 1,966                         & 100,000                                 & 590                      \\ \hline
DDoS                         & \multirow{7}{*}{DoS}                          & \multirow{7}{*}{380,699}      & \multirow{7}{*}{266,489}                & \multirow{7}{*}{114,210} \\ \cline{1-1}
DoS   GoldenEye              &                                               &                               &                                         &                          \\ \cline{1-1}
DoS   Hulk                   &                                               &                               &                                         &                          \\ \cline{1-1}
DoS   Slow-httptest          &                                               &                               &                                         &                          \\ \cline{1-1}
DoS   Slowloris              &                                               &                               &                                         &                          \\ \cline{1-1}
Heartbleed                   &                                               &                               &                                         &                          \\ \hline
Port-Scan                    & Sniffing                                      & 158,930                       & 111,251                                 & 47,679                   \\ \hline
SSH-Patator                  & \multirow{2}{*}{Brute-Force}                  & \multirow{2}{*}{13,835}       & \multirow{2}{*}{100,000}                & \multirow{2}{*}{4,150}   \\ \cline{1-1}
FTP-Patator                  &                                               &                               &                                         &                          \\ \hline
Infiltration                 & Infiltration                                  & 36                            & 100,000                                 & 11                       \\ \hline
Web   Attack – Brute Force   & \multirow{4}{*}{Web Attack}                   & \multirow{4}{*}{2,180}        & \multirow{4}{*}{100,000}                & \multirow{4}{*}{654}     \\ \cline{1-1}
Web   Attack – Sql Injection &                                               &                               &                                         &                          \\ \cline{1-1}
Web   Attack – XSS           &                                               &                               &                                         &                          \\ \hline
\end{tabular}%
}
\label{CIC}%
\end{table}%

\subsection{Performance Analysis of Known Intrusion Detection}
To evaluate the proposed IDS for known intrusion detection, the ML models in the signature-based IDS are trained and tested on the two labeled datasets that represent intra-vehicle and external vehicular network traffic data. The results of 10-fold cross-validation of the proposed models on the CAN-intrusion dataset \cite{GIDS} and the CICIDS2017 dataset \cite{CIC} are shown in Tables \ref{CAN1} and \ref{CIC1}, respectively.

\begin{table}[]
\centering%
\caption{Performance Evaluation of Classifiers on The CAN-Intrusion-Dataset}
\setlength\extrarowheight{1pt}
\scalebox{0.85}{
\begin{tabular}{|>{\centering\arraybackslash}m{9em}|>{\centering\arraybackslash}m{3em}|>{\centering\arraybackslash}m{3em}|>{\centering\arraybackslash}m{3em}|>{\centering\arraybackslash}m{3em}|>{\centering\arraybackslash}m{4em}|}
\hline
\textbf{Method}    & \textbf{Acc (\%)} & \textbf{DR (\%)} & \textbf{FAR (\%)} & \textbf{F1} & \textbf{Execution Time (S)} \\ \hline
KNN \cite{known1}       & 97.4              & 96.3             & 5.3               & 0.934       & 911.6                       \\ \hline
SVM \cite{known1}       & 96.5              & 95.7             & 4.8               & 0.933       & 13765.6                     \\ \hline
XYF-K \cite{can-mdpi}      & 99.1                 & 98.39            & 0.0                & 0.9879        & -                           \\ \hline
SAIDuCANT \cite{reco1}      & 87.21                 & 86.66            & 1.76                & 0.92        & -                           \\ \hline
SSAE \cite{known3}      & -                 & 98.5             & 2.0                 & 0.98        & -                           \\ \hline
DCNN \cite{known4}      & 99.93                 & 99.84             & 0.16                 & 0.9991        & -                           \\ \hline
LSTM-Autoencoder \cite{tits}      & 99.0                 & 99.0            & 0.0                & 0.99        & -                           \\ \hline
\textbf{MTH-IDS}    & \textbf{99.999}            & \textbf{99.999}           & \textbf{0.0006}            & \textbf{0.99999}     & \textbf{365.3}                       \\ \hline
\end{tabular}
}
\label{CAN1}
\end{table}

\begin{table}[]
\centering%
\caption{Performance Evaluation of Classifiers on The CICIDS2017 Dataset}
\setlength\extrarowheight{1pt}
\scalebox{0.85}{
\begin{tabular}{|>{\centering\arraybackslash}m{8em}|>{\centering\arraybackslash}m{3em}|>{\centering\arraybackslash}m{3em}|>{\centering\arraybackslash}m{3em}|>{\centering\arraybackslash}m{3em}|>{\centering\arraybackslash}m{4em}|}
\hline
\textbf{Method}    & \textbf{Acc (\%)} & \textbf{DR (\%)} & \textbf{FAR (\%)} & \textbf{F1} & \textbf{Execution Time (S)} \\ \hline
KNN \cite{CIC}              & 96.3     & 96.2    & 6.3      & 0.963   & 15243.6            \\\hline
RF \cite{CIC}              & 98.82    & 98.8    & 0.145    & 0.988   & 1848.3             \\\hline
SU-IDS \cite{known3c}      & 99.13     & 99.65    & 1.4      & -       & -                  \\\hline
STDeepGraph \cite{known4c}      & 99.4     & 98.6    & 1.3      & -       & -                  \\\hline
DBN \cite{knowncim1}            & 98.95    & 95.82   & 4.19    & 0.9581   & -            \\\hline
GAN-RF \cite{knowncim2}            & 99.83    & 98.68   & 7.24     & 0.9504   & -            \\\hline
DeepCoin \cite{knowncim3}            & 99.811    & 94.10   & 0.986     & -   & -            \\\hline
Multi-SVM \cite{known1c}            & 98.55    & 98.22   & 0.41     & 0.983   & 34896.5            \\\hline
PCA-RF \cite{knowncn2}            & 99.6    & 99.6   & 1.0     & 0.996   & -            \\\hline
MTH-IDS (Without FS \& HPO)                 & 99.861   & 99.753  & 0.110    & 0.99860 & 5238.4             \\\hline
\textbf{MTH-IDS (Multi-Class Model) }         & \textbf{99.879}   & \textbf{99.818}  & \textbf{0.101}    & \textbf{0.99879} & \textbf{1563.4}            \\\hline
\textbf{MTH-IDS (Binary Model)} & \textbf{99.895}   & \textbf{99.806}  & \textbf{0.084}    & \textbf{0.99895} & \textbf{478.2}      \\\hline
\end{tabular}
}
\label{CIC1}
\end{table}

As shown in Table \ref{CAN1}, after utilizing the proposed IG-FCBF feature selection method with the optimized correlation threshold ($\alpha=0.9$) to select the top four significant features (“CAN ID”, “DATA[5]”, “DATA[3]”, and “DATA[1]”), the proposed signature-based IDS can reach high accuracy of 99.999\% on the CAN-intrusion dataset. The proposed method is compared with recent promising approaches proposed in the literature \cite{known1}–\cite{reco1}, \cite{known3}–\cite{tits}, as shown in Table \ref{CAN1}. As the attack and normal patterns in this dataset can be obviously distinguished, most of the compared methods also achieve high accuracy.
The authors in \cite{known4} proposed a deep convolutional neural network (DCNN) method that can achieve a high average F1-score of 0.9991, but it requires the model training machines to have graphics processing units (GPUs), making it difficult to be used in vehicle-level systems due to budget constraints. 
For six other compared approaches proposed in the recent literature, our proposed system achieves at least 0.89\% accuracy and F1-score improvement. 

The experimental results on the CICIDS2017 dataset are shown in Table \ref{CIC1}. By implementing the proposed IG-FCBF method with the optimized correlation threshold ($\alpha=0.9$) to select 20 features from 80 original features and using HPO methods to obtain optimized ML models, the F1-score of the proposed IDS has improved from 99.861\% to 99.879\%, and the execution time has decreased by 70.2\%, as shown in Table \ref{CIC1}. This justifies the proposed FS method and the BO-TPE method can greatly improve the system efficiency and slightly improve the model accuracy.
In addition to the multi-classification results used to evaluate the IDS’s capacity to detect various types of attacks, the IDS is also implemented to train a binary classification model that can distinguish between normal and abnormal network traffic data and return one of these two labels. As shown in Table  \ref{CIC1}, the proposed IDS reaches 99.895\% accuracy and saves 69.4\% of the execution time by training the binary classification model. Binary classifiers and multi-classifiers can be chosen according to the specific needs of users.

Although there are many existing works that evaluate their models on the CICIDS2017 dataset, most of them have different or inexplicit validation configurations. Nevertheless, the proposed model is quantitatively compared with recent promising methods that achieve good performance on the CICIDS2017 dataset and have a similar validation configuration, proposed in the literature \cite{CIC}, \cite{known3c}-\cite{known4c}, \cite{knowncim1}-\cite{knowncn2}. 
The approaches proposed in the literature \cite{knowncim1}-\cite{knowncim3} have achieved high accuracy (98.95\%-99.83\%), but a relatively low detection rate (94.10\%-98.68\%) and F1-score (0.9504-0.9581) due to the imbalanced dataset. Other methods proposed in the literature \cite{CIC}, \cite{known3c}-\cite{known4c}, \cite{known1c}-\cite{knowncn2} have also achieved high accuracy or F1-scores, but still slightly lower than our proposed model.
To summarize, even though the comparison with existing approaches is not a straightforward process, the proposed system shows better performance by at least achieving a 0.279\% higher F1-score than of the same-level validation configuration approaches by quantitative comparison.

Therefore, the experimental results show that the proposed IDS can efficiently separate normal and malicious network traffic data and effectively detect various types of known cyber-attacks in vehicle systems. 
\subsection{Performance Analysis of Unknown Intrusion Detection}

At this stage, all the models in the anomaly-based IDS are trained for binary classification by labeling the instances of all attack types as “attack” and normal instances as “normal”. In the proposed system, after being evaluated by the signature-based IDS, all data samples of known attack types will be returned, and other normal instances will be labeled “suspicious” and passed to the anomaly-based IDS to determine whether any unknown attacks exist.

For the CAN-intrusion-dataset that represents intra-vehicle network traffic data, each type of CAN attack is regarded as a new attack type in each experiment. The evaluation results of unknown CAN attack detection are shown in Table \ref{CAN2}. For DoS, gear spoofing, and RPM spoofing attack types, the proposed system can reach 100\% detection rates, very low false alarm rates (0.0\%-0.449\%), and very high F1-scores (more than 0.997). However, the DR and F1 for the fuzzy attack are much lower (73.053\% and 0.84389). This is because the feature values of the fuzzy attack packets are random numerical values, and certain random values can be very similar to normal packets, making it difficult for unsupervised learning algorithms to distinguish them. Moreover, the performance of the proposed anomaly-based IDS is compared to the CL-k-means model without biased classifiers. Table \ref{CAN2} shows that the F1 score of the proposed MTH-IDS on the CAN-intrusion-dataset can be largely improved from 0.82643 to 0.96307 by implementing the two biased classifiers after the CL-k-means model. 
To summarize, the proposed system can effectively detect most unknown attacks on intra-vehicle networks except fuzzy attacks.

\begin{table}[]
\centering%
\caption{Performance Evaluation on Each Type of Unknown Attack of The CAN-Intrusion-Dataset}
\setlength\extrarowheight{1pt}
\scalebox{0.85}{\begin{tabular}{|>{\centering\arraybackslash}m{10em}|>{\centering\arraybackslash}m{4.5em}|>{\centering\arraybackslash}m{3.5em}|>{\centering\arraybackslash}m{4em}|>{\centering\arraybackslash}m{3.5em}|}
\hline
\textbf{Attack Type}       & \textbf{Validation Instances} & \textbf{DR (\%)} & \textbf{FAR (\%)} & \textbf{F1} \\ \hline
DoS        & 1,289,386                     & 100.0              & 0.0            & 1.0     \\ \hline
Fuzzy         & 1,193,712                     & 73.053              & 0.057             & 0.84389     \\ \hline
Gear Spoofing        & 1,299,117                     & 100.0              & 0.449             & 0.99736     \\ \hline
RPM Spoofing         & 1,356,762                     & 100.0              & 0.003             & 0.99998     \\ \hline
\textbf{Average (MTH-IDS)}         & \textbf{5,138,977 }                  &  \textbf{93.740}	& \textbf{0.128}	& \textbf{0.96307}     \\ \hline
Average (CL-k-means)         & 5,138,977                     &79.233	&0.261	&0.82643     \\ \hline
\end{tabular}
}
\label{CAN2}
\end{table}

On the other hand, training on more different types of attack samples enables us to design a more comprehensive IDS that can detect more unknown attack types effectively \cite{moretypes}. Therefore, several experiments were conducted on the CICIDS2017 dataset that contains data samples of 14 different common cyber-attacks types to illustrate potential attacks launched on external vehicular networks. In each experiment of the validation process, each type of attack is regarded as an unknown attack, and the results are shown in Table \ref{CIC2}.

\begin{table}[]
\centering%
\caption{Performance Evaluation on Each Type of Unknown Attack of The CICIDS2017 Dataset}
\setlength\extrarowheight{1pt}
\scalebox{0.85}{\begin{tabular}{|>{\centering\arraybackslash}m{10em}|>{\centering\arraybackslash}m{4.5em}|>{\centering\arraybackslash}m{3.5em}|>{\centering\arraybackslash}m{4em}|>{\centering\arraybackslash}m{3.5em}|}
\hline
\textbf{Attack Type}       & \textbf{Validation Instances} & \textbf{DR (\%)} & \textbf{FAR (\%)} & \textbf{F1} \\ \hline
Bot                        & 3,932                         & 63.276           & 21.669            & 0.68426     \\ \hline
DDoS                       & 256,054                       & 62.697           & 11.698            & 0.71902     \\ \hline
DoS GoldenEye              & 20,586                        & 83.931           & 20.461            & 0.82127     \\ \hline
DoS Hulk                   & 462,146                       & 67.440            & 11.806            & 0.75248     \\ \hline
DoS Slow-httptest          & 10,998                        & 76.687           & 19.094            & 0.78339     \\ \hline
DoS Slowloris              & 11,592                        & 83.834           & 7.902             & 0.87447     \\ \hline
FTP-Patator                & 15,876                        & 51.298           & 12.686            & 0.62564     \\ \hline
Heartbleed                 & 22                            & 100.0              & 18.182            & 0.91667     \\ \hline
Infiltration               & 72                            & 72.222           & 5.556             & 0.81250     \\ \hline
Port-Scan                  & 317,860                       & 98.962           & 17.849            & 0.91288     \\ \hline
SSH-Patator                & 11,794                        & 95.828           & 23.351            & 0.87443     \\ \hline
Web Attack – Brute Force   & 3,014                         & 89.516           & 17.319            & 0.86558     \\ \hline
Web Attack – Sql Injection & 42                            & 95.238           & 23.810            & 0.86957     \\ \hline
Web Attack – XSS           & 1,304                         & 3.681            & 14.417            & 0.06233     \\ \hline
\textbf{Average (MTH-IDS)}                    & \textbf{1,115,292}                     & \textbf{75.943}           & \textbf{13.882}            & \textbf{0.80013}     \\ \hline
Average (CL-k-means)         & 1,115,292                     & 72.682              & 15.357             & 0.77305     \\ \hline
\end{tabular}
}
\label{CIC2}
\end{table}

From Table \ref{CIC2}, it can be seen that the proposed system exhibits different performances when applied to the experiments of different types of unknown attacks. By implementing the proposed methods, the false alarm rates for most of the attack types are at a low level of less than 20\%. The detection rates for the “Heartbleed”, “Port-Scan”, “SSH-Patator”, “Web Attack – Brute Force”, and “Web Attack – Sql Injection” attacks are high (from 89.516\% to 100\%), while the detection rates for other types of attacks are relatively lower (from 51.298\% to 83.931\%). The F1-scores for most of the attack types are larger than 0.80. The only type of attack that the proposed system cannot detect effectively is the “Web Attack – XSS” whose results show a very low F1-score (0.062), because their data distribution is very similar to normal data distributions. The average F1-score of the proposed MTH-IDS on all the attacks is 0.80013, which is higher than the CL-k-means model without biased classifiers (0.77305).

Thus, the proposed IDS can detect most of the previously-unseen types of attacks with a relatively high detection rate and a relatively low false alarm rate on both intra-vehicle and external vehicular networks. Nevertheless, there is still some room for improvement since effectively detecting zero-day attacks is still an unsolved research problem.

\subsection{Vehicle-Level Model Evaluation and Discussion}

\begin{table}[]
\centering%
\caption{Performance Evaluation on The Untouched Test Set}
\setlength\extrarowheight{1pt}
\scalebox{0.85}{\begin{tabular}{|>{\centering\arraybackslash}m{9.5em}|>{\centering\arraybackslash}m{4.5em}|>{\centering\arraybackslash}m{4em}|>{\centering\arraybackslash}m{4em}|>{\centering\arraybackslash}m{4em}|}
\hline
\textbf{Dataset}       & \textbf{Acc (\%)} & \textbf{DR (\%)} & \textbf{FAR (\%)} & \textbf{F1} \\ \hline
CAN-intrusion-dataset                        & 99.99                         & 100.0           & 0.00005            & 0.9999     \\ \hline
CICIDS2017                       & 99.88                       & 99.77           & 0.10            & 0.9988     \\ \hline
\end{tabular}
}
\label{test}
\end{table}

\begin{table}[]
\centering%
\caption{Model Evaluation on A Vehicle-Level System}
\setlength\extrarowheight{1pt}
\scalebox{0.85}{
\begin{tabular}{|>{\centering\arraybackslash}m{5.5em}|>{\centering\arraybackslash}m{5em}|>{\centering\arraybackslash}m{5em}|>{\centering\arraybackslash}m{5.5em}|>{\centering\arraybackslash}m{5.5em}|}
\hline
\textbf{System Component}          & \textbf{Dataset 1: Avg Test Time (ms)} & \textbf{Dataset 2: Avg Test Time (ms)} & \textbf{Dataset 1: Model Space (MB)} & \textbf{Dataset 2: Model Space (MB)}  \\ 
\hline
Z-score      & 0.028                          & 0.031                          & -                            & -                             \\ 
\hline
KPCA                & 0.005                          & 0.009                          & 0.002                      & 0.006                        \\ 
\hline
Stacking           & 0.389                          & 0.297                          & 2.02                       & 14.36                        \\ 
\hline
CL-k-means         & 0.145                          & 0.157                          & 0.48                        & 0.78                        \\ 
\hline
Biased Classifiers & 0.007                          & 0.015                          & 0.11                       & 1.06                        \\ 
\hline
\textbf{Sum}                & \textbf{0.574}                          & \textbf{0.509}                          & \textbf{2.61}                       & \textbf{16.21}                       \\
\hline
\end{tabular}
}
\label{realtime}
\end{table}

\begin{figure}
     \centering
     \includegraphics[width=7.5cm]{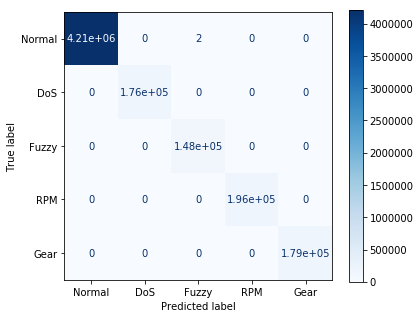}
     \caption{Confusion matrix for the test set of CAN-intrusion-dataset.} \label{cm1}
\end{figure}
\begin{figure}
     \centering
     \includegraphics[width=8.5cm]{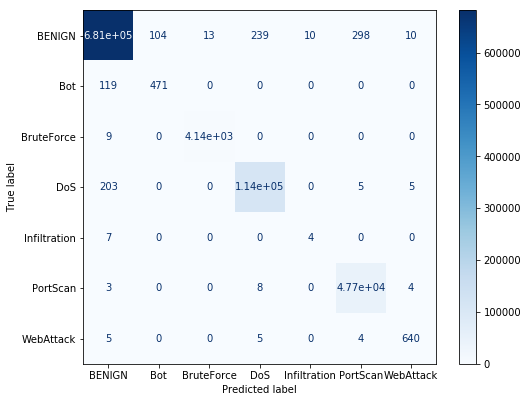}
     \caption{Confusion matrix for the test set of CICIDS2017 dataset.} \label{cm2}
\end{figure}

The proposed IDS with all trained models are tested on Raspberry Pi 3, a vehicle-level machine, to evaluate its feasibility in vehicular environments. 
Moreover, implementing the proposed model on the untouched test sets can evaluate its generalizability.

The experimental results on the test sets are shown in Table \ref{test}. As shown in Table \ref{test}, the F1-scores of the proposed IDS on the 30\% test sets of the CAN-intrusion-dataset and CICIDS2017 dataset are 99.99\% and 99.88\%, respectively. Moreover, the confusion matrices of evaluating the proposed method on the test sets of the CAN-intrusion-dataset and CICIDS2017 dataset are shown in Fig. \ref{cm1} and Fig. \ref{cm2}, respectively. For the CAN-intrusion-dataset, as shown in Fig. \ref{cm1}, the proposed method can accurately detect all the DoS, RPM spoofing, and gear spoofing attack samples, and only has two false alarms for the fuzzy attack detection. These results are in line with other similar works from the literature that used the same dataset. For example, the authors in \cite{known4} also achieved high accuracy of 99.93\%. The main reason for achieving high accuracy is that the large difference between the attack and normal patterns in the CAN-intrusion-dataset can be obviously distinguished. For the CICIDS2017 results shown in Table \ref{test} and Fig. \ref{cm2}, the attack patterns are more difficult to be distinguished than the CAN-intrusion-dataset, but the classification error rate is still at a very low level (0.12\%), except for the infiltration attack type that has 7 misclassified samples out of 11 test samples. This is because the number of data samples for the infiltration attack is only 36, which is insufficient to train an effective classifier to accurately identify this attack. Nevertheless, the proposed model can accurately detect other attacks with an overall F1-score of 99.88\%.
As the test sets were untouched before the hold-out validation, the high performance indicates the strong generalizability of the proposed framework on new datasets.

The proposed model can achieve high performance without over-fitting mainly due to the following reasons \cite{overfitting}: 
\begin{enumerate}
\item It trains on large-sized datasets, as using more data samples can improve the generalizability of the proposed method.
\item It implements a comprehensive feature engineering method to improve the generalizability by removing irrelevant and misleading features that may cause over-fitting.
\item It uses the stacking ensemble method to combine the results of base learners. Ensemble models often have better generalizability than single models because combining the single learners can reduce the variance of estimation and prevent over-fitting.
\end{enumerate}

On the other hand, to deploy the proposed IDS in real-world vehicle systems, the real-time requirements of vehicle safety services should be met. For each packet transmitted on vehicular networks, the alarm generation time should be less than 10ms to meet the real-time requirements, as described in Section III-D. For each packet passed to the proposed IDS, it will be processed by Z-score normalization and four trained models: KPCA, stacking, CL-k-means, and a biased classifier (either FN-based or FP-based). As shown in Table \ref{realtime}, the average of the processing time for each packet in the CAN-intrusion-dataset (Dataset 1) and the CICIDS2017 dataset (Dataset 2) is only 0.574 ms and 0.509 ms, respectively, which is much lower than the vehicular network security latency requirement (10ms). Moreover, since we trained the models on the small-size subsets obtained by the proposed k-means cluster sampling method, the total size of the trained models for the intra-vehicle network IDS and external network IDS is only 2.61 MB and 16.21 MB, respectively, which is much less than the memory limit of vehicle-level machines that often have more than 1 GB RAM, like Raspberry Pi 3. Thus, the experimental results show the feasibility of applying the proposed system to real-time vehicle systems.

\section{Conclusion}
To enhance IoV security, this work proposed a multi-tiered hybrid intrusion detection system (MTH-IDS) model that can detect various types of known and zero-day cyber-attacks on both intra-vehicle and external-vehicular networks for modern vehicles. The proposed MTH-IDS consists of two traditional ML stages (data pre-processing and feature engineering) and four main tiers of learners utilizing multiple machine learning algorithms. Through data pre-processing and feature engineering, the quality of the input data can be significantly improved for more accurate model learning. The first tier of the proposed system consists of four tree-based supervised learners used for known attack detection, while the second tier comprises the BO-TPE and stacking models for supervised base learner optimization to achieve higher accuracy. The third tier consists of a novel CL-k-means unsupervised model used for unknown/zero-day attack detection. Lastly, BO-GP and two biased classifiers are used to construct the fourth tier for unsupervised learner optimization. The four tiers of learning models enable the proposed MTH-IDS to achieve optimal performance for both known and unknown attack detection in vehicular networks.

Through the performance evaluation of the proposed IDS on the two public datasets that represent intra-vehicle and external vehicular network data, the proposed system can effectively detect various types of known attacks with accuracies of 99.99\% and 99.88\% on the CAN-intrusion-dataset and CICIDS2017 dataset, respectively. Moreover, the proposed system can detect various types of unknown attacks with average F1-scores of 0.963 and 0.800 on the CAN-intrusion-dataset and CICIDS2017 dataset, respectively. The experimental results on a vehicle-level machine also show the feasibility of the proposed system in real-time environments. In future work, the proposed anomaly-based IDS framework can be further improved by doing research on other unsupervised learning and online learning methods.


\ifCLASSOPTIONcaptionsoff
  \newpage
\fi


\begin{IEEEbiography}[{\includegraphics[width=1in,height=1.25in,clip,keepaspectratio]{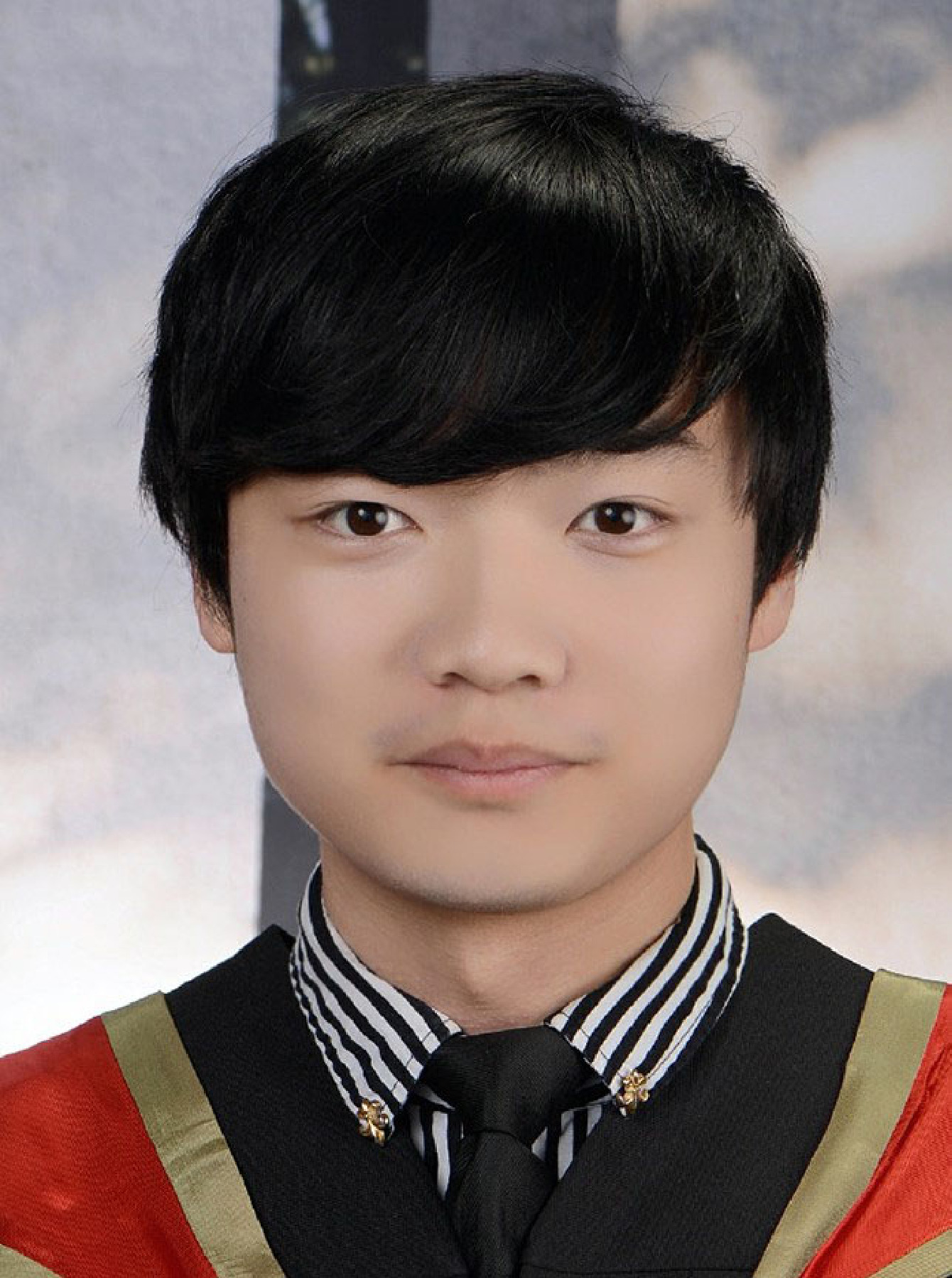}}]{Li Yang}received the B.E. degree in computer science from Wuhan University of Science and Technology, Wuhan, China in 2016 and the MASc degree in Engineering from University of Guelph, Guelph, Canada, 2018. Since 2018 he has been working toward the Ph.D. degree in the Department of Electrical and Computer Engineering, Western University, London, Canada. His research interests include cybersecurity, machine learning, time-series data analytics, and intelligent transportation systems.
\end{IEEEbiography}

\begin{IEEEbiography}[{\includegraphics[width=1in,height=1.25in,clip,keepaspectratio]{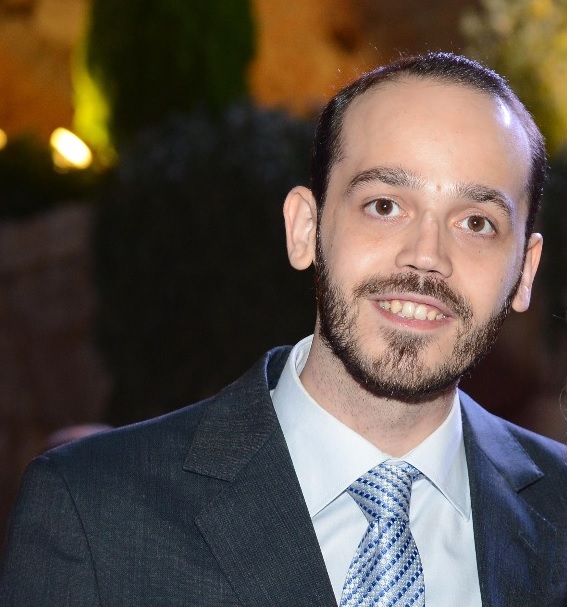}}] {Abdallah Moubayed}received his Ph.D. in Electrical \& Computer Engineering from the University of Western Ontario in August 2018,  his M.Sc. degree in Electrical Engineering from King Abdullah University of Science and Technology, Thuwal, Saudi Arabia in 2014, and his B.E. degree in Electrical Engineering from the Lebanese American University, Beirut, Lebanon in 2012. Currently, he is a Postdoctoral Associate in the Optimized Computing and Communications (OC2) lab at University of Western Ontario.  His research interests include wireless communication, resource allocation, wireless network virtualization, performance \& optimization modeling, machine learning \& data analytics, computer network security, cloud computing, and e-learning.
\end{IEEEbiography}

\begin{IEEEbiography}[{\includegraphics[width=1in,height=1.25in,clip,keepaspectratio]{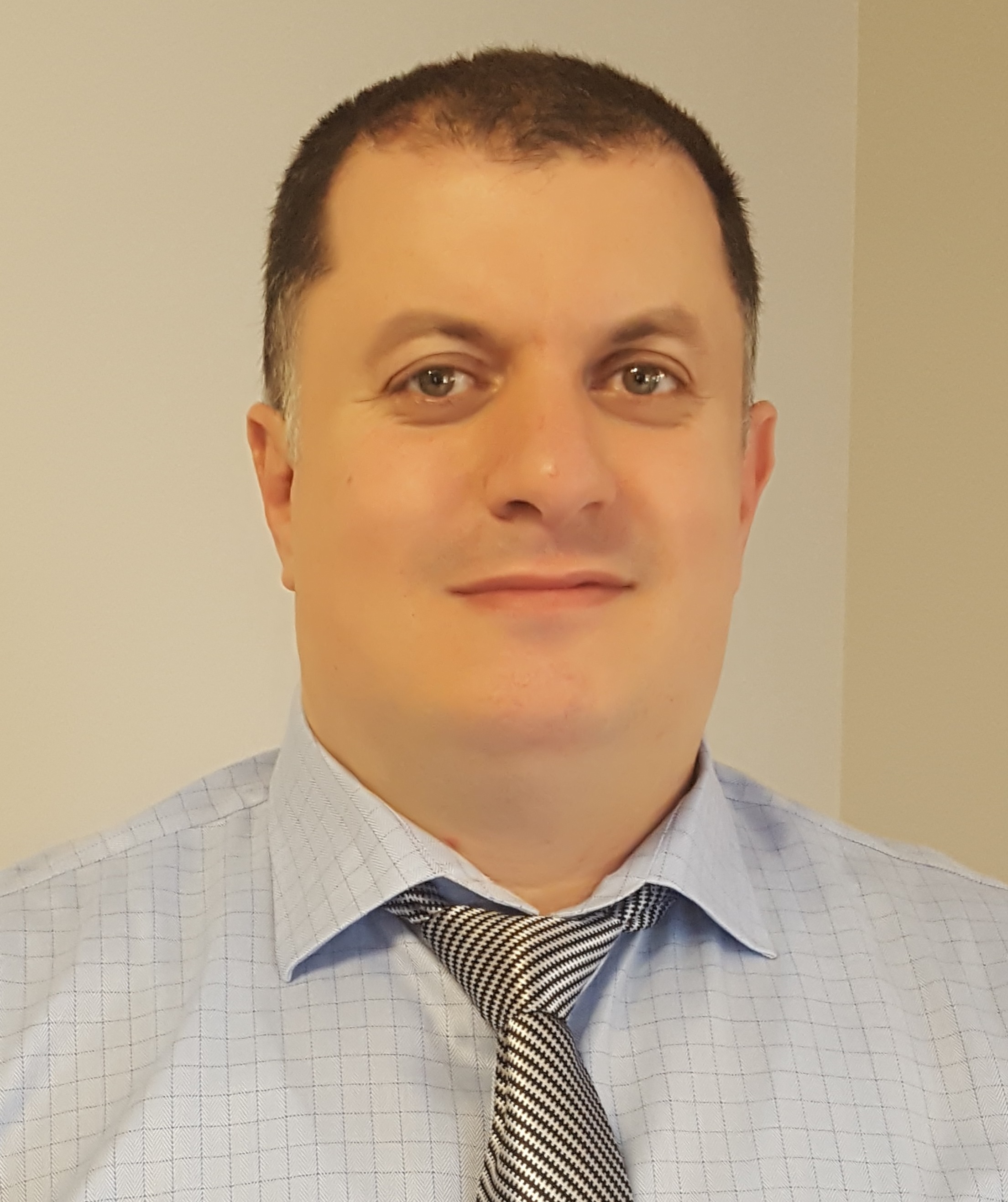}}] {Abdallah Shami}is a professor with the ECE Department at Western University, Ontario, Canada. He is the Director of the Optimized Computing and Communications Laboratory at Western University (https://www.eng.uwo.ca/oc2/). He is currently an associate editor for IEEE Transactions on Mobile Computing, IEEE Network, and IEEE Communications Surveys and Tutorials. He has chaired key symposia for IEEE GLOBECOM, IEEE ICC, IEEE ICNC, and ICCIT. He was the elected Chair of the IEEE Communications Society Technical Committee on Communications Software (2016-2017) and the IEEE London Ontario Section Chair (2016-2018).
\end{IEEEbiography}


\end{document}